\documentclass{article}

\usepackage{arxiv}
\usepackage[numbers,sort&compress]{natbib}
\usepackage{chemformula} 
\usepackage[T1]{fontenc} 
\usepackage{graphicx}
\usepackage{subcaption}
\usepackage{cleveref}
\usepackage{color}
\usepackage{adjustbox}
\usepackage{makecell}
\usepackage{soul}
\usepackage{booktabs}
\usepackage{import}

\title{Atmospheric water vapor condensation on engineered interfaces: Busting the myths}

\author{Tibin M. Thomas\\
	Department of Mechanical Engineering\\
	Indian Institute of Technology Madras\\
	Chennai 600036, India \\
	\texttt{me17d042@smail.iitm.ac.in} \\
	\And
	Pallab Sinha Mahapatra \\
	Department of Mechanical Engineering\\
	Indian Institute of Technology Madras\\
	Chennai 600036, India \\
	\texttt{pallab@iitm.ac.in} \\
	\AND
	Ranjan Ganguly \\
	Department of Power Engineering\\
	Jadavpur University\\
	Kolkata 700106, India \\
	\texttt{ranjan.ganguly@jadavpuruniversity.in}\\
 	\And
 	Manish K. Tiwari \\
 	Nanoengineered Systems Laboratory;\\
 	Wellcome/EPSRC Centre for\\
 	Interventional and Surgical Sciences\\
	UCL, UK\\
 	\texttt{m.tiwari@ucl.ac.uk}
}
\renewcommand{\shorttitle}{Atmospheric water vapor condensation}

\begin{document}
\maketitle

\begin{abstract}
Condensing atmospheric water vapor on surfaces is a sustainable approach to potentially address the potable water crisis. However, despite extensive research, a key question remains: what is the physical mechanism governing the condensation from humid air and how significantly does it differ from pure steam condensation? The answer may help define an optimal combination of the mode and mechanism of condensation as well as the surface wettability for best possible water harvesting efficacy. Here we show that this lack of clarity is due to the differences in heat transfer characteristics during condensation from pure vapor and humid air environments. Specifically, during condensation from humid air, the thermal resistance across the condensate is non-dominant and the energy transfer is controlled by vapor diffusion and condensate drainage. This leads to filmwise condensation on superhydrophilic surfaces, offering the highest water collection efficiency. To demonstrate this, we measured condensation rate on different sets of superhydrophilic and superhydrophobic surfaces in a wide degree of subcooling (10 - 26 $^\circ C$) and humidity-ratio differences (5 - 45 \textit{g/kg of dry air}). The resulting condensation rate is enhanced by 57 - 333 \% on the superhydrophilic surfaces as compared to the superhydrophobic ones. The findings of this study challenges the nearly century-old scientific ambiguity about the mechanism of vapor condensation from humid air. Our findings will lead to the design of efficient atmospheric water harvesting systems.
\end{abstract}


\section{INTRODUCTION}
The dwindling reserve of potable water on the planet as a result of industrialization, climate change, and environmental pollution is a major challenge in the coming decades. Recently, it has been estimated that half a billion of the global population is facing severe water scarcity throughout the year, while about four billion people are facing the same for at least one month of the year \cite{mekonnen2016four}. Capturing water vapor from the humid air by condensation is a sustainable water-production technology, particularly in arid environments. \cite{tu2018progress, nagar2020clean, haechler2021exploiting}. Condensing water from the atmosphere is an attractive means to address this potable water shortage. In the earth's atmosphere, the mass fraction of water vapor is always below 5 \%, with more than 95 \% being non-condensable gases (NCG). Condensation of the vapor requires a subcooled surface for both pure vapor and vapor in humid air. It begins with the formation of liquid nuclei on the surface once its temperature falls below the saturation temperature. The mechanism of vapor condensation in the presence of NCG is distinct than pure vapor condensation due to the formation of thermal and concentration boundary layers outside the liquid-vapor interface, which are not observed during pure vapor condensation. Along with the growth of liquid nuclei, the NCG molecules present in the humid air are accumulated over the interface of the nuclei and form an NCG diffusion layer \cite{kroger1968condensation, zheng2018modeling}. This layer acts as a barrier to water vapor diffusion from the bulk humid air to the interface, and even a small quantity of NCG (mass fraction of around 0.5 \%) can reduce heat transfer rates by 50 \% when compared to a pure vapor environment \cite{othmer1929condensation, slegers1970laminar}. Furthermore, depending on the water vapour concentration, the condensation heat transfer rate in humid air can be 2 to 3 orders of magnitude lower than that in pure vapour \cite{oestreich2019experimental,thomas2021condensation}. Prior research over the past century has shown that the condensation heat transfer rate of the dropwise condensation (DWC) mode can be one order of magnitude higher than the filmwise condensation (FWC) mode under pure vapor conditions \cite{schmidt1930condensation, le1965experimental, cha2020dropwise}. However, it remains unclear which mode of condensation is better for water harvesting from humid air, despite decades of research. 
\begin{table}
\centering
\resizebox{\textwidth}{!}{
\begin{tabular}{ccccccc}
\toprule
Ref   & \begin{tabular}[c]{@{}c@{}}Substrate\\type\end{tabular} & \begin{tabular}[c]{@{}c@{}}Surface \\morphology\end{tabular}& \begin{tabular}[c]{@{}c@{}}Environmental\\ condition\end{tabular} & \begin{tabular}[c]{@{}c@{}}Subcooling\\ $\Delta T$ ($^\circ C$) \end{tabular} & \begin{tabular}[c]{@{}c@{}}Difference in \\ humidity ratio \\ $\Delta \omega$ ($g/kg$)\end{tabular} & \begin{tabular}[c]{@{}c@{}}Percentage \\ change in \\ FWC mass flux \\with DWC (\%) \end{tabular} \\ \midrule 
\cite{choo2015water} & \makecell{Silicon\\substrate} & \makecell{$TiO_2$\\nanorods} & T=25 $^\circ C$, RH=93.5 \%                                                     & 20.87      & 14.06                                                                   & +109                                                                                 \\
\cite{mahapatra2016key} & \makecell{Aluminum\\substrate} & \makecell{Hierarchical\\$Al_2O_3$} & T=20 $^\circ C$, RH=80 \%                                                      & 14.15      & 7.24                                                                    & +50                                                                                  \\
\cite{mahapatra2016key} & \makecell{Aluminum\\substrate} & \makecell{Hierarchical\\$Al_2O_3$} & T=35 $^\circ C$, RH=80 \%                                                      & 8.6        & 11.84                                                                   & +37.5                                                                                \\
\cite{seo2016effects} & \makecell{Copper tube} & \makecell{$CuO$\\nanostructures} & T=40 $^\circ C$, RH=80 \%                                                      & 32.5       & 33.69                                                                   & +4.5                                                                                 \\
\cite{hou2020tunable} & \makecell{Copper foil} & \makecell{$CuO$\\nanostructures} & T=26 $^\circ C$, RH=50 \%                                                      & 12.8       & 6.13                                                                    & +26.4                                                                                \\
\cite{pinheiro2019water} & \makecell{Steel tube} & \makecell{Carbon\\nanotubes} & T=40 $^\circ C$, RH=80 \%                                                      & 31.9       & 33.48                                                                   & -29.4                                                                                \\
\cite{pinheiro2020vertically} & \makecell{Steel tube} & \makecell{Carbon\\nanotubes} & T=30 $^\circ C$, RH=80 \%                                                       & 14.2       & 12.85                                                                   & -34.4                                                                                \\
\cite{gupta2019background} & \makecell{Aluminum\\substrate} & \makecell{Hierarchical\\$Al_2O_3$} & T=24.5 $^\circ C$, RH=70 \%                                                     & 13.7       & 8.11                                                                    & +21                                                                                  \\
\cite{gupta2019background} & \makecell{Aluminum\\substrate} & \makecell{Hierarchical\\$Al_2O_3$} & T=24.5 $^\circ C$, RH=80 \%                                                    & 15.8       & 10.08                                                                   & -1.2                                                                                 \\
\cite{gupta2019background} & \makecell{Aluminum\\substrate} & \makecell{Hierarchical\\$Al_2O_3$} & T=24.5 $^\circ C$, RH=90 \%                                                    & 17.7       & 12.07                                                                   & -27.8                                                                                \\
\cite{lee2020improved} & \makecell{Aluminum\\substrate} & \makecell{Hierarchical\\$Al_2O_3$} & T=14 $^\circ C$, RH=100 \%                                                     & 9          & 4.57                                                                    & +78.5 \\                                                                              
\cite{lee2012water} & \makecell{Silicon\\substrate} & \makecell{Rough\\nanostructures} & T=25 $^\circ C$, RH=92.5 \%                                                     & 13.69          & 10.91                                                                    & -12.1  \\
\cite{nagar2021scalable} & \makecell{Aluminum\\substrate} & \makecell{Hierarchical\\$Al_2O_3$} & T=23.6 $^\circ C$, RH=50 \%                                                     & 8          & 5.25                                                                    & +8.9
\\ \bottomrule
\end{tabular}}

\caption{\textbf{Summary of previous studies reporting the overall mass flux during condensation on sub-cooled superhydrophobic and superhydrophilic surfaces in vertical orientation.} The reported results are too dispersed and cannot be correlated with operating conditions such as degree of subcooling or humidity ratio difference. The influence of several experimental uncertainties, such as dynamic changes in relative humidity within the condensation chamber, perturbation or velocity of the condensing fluid, edge effects of the condensing surface, surface aspect ratio, and so on, could be implicated for the inconclusive trend. The compiled data reveals a lack of understanding regarding the overall impact of wettability modification on water yield during humid air condensation.}
\label{tab1}
\end{table}

Over the last two decades, advances in surface chemistry and micro/nano fabrication techniques have inspired researchers to custom-engineer the wettability of condensing surface for better condensation effectiveness, which is required for water-energy nexus applications \cite{sharma2017growth, chen2015exploiting, rykaczewski2013multimode, olceroglu2016self, orejon2019dropwise, winter2020dewetting, zhao2015condensate, sharma2019self}. Dropwise condensation on a superhydrophobic surface has been proposed for water harvesting in several publications due to the early departure or enhanced mobility of condensing droplets, allowing greater access to the dry condensing area for continuous nucleation. In addition, filmwise condensation on superhydrophilic surfaces has also been proposed for water harvesting. Table \ref{tab1} summarises the research articles that reported the overall condensation rate on the non-wetting superhydrophobic and wetting superhydrophilic surfaces in the atmospheric humid air conditions. Clearly, the majority of the works cited in Table \ref{tab1} report that filmwise mode is more effective than dropwise mode for humidity harvesting. However, these observations did not receive comprehensive attention and, as a consequence, has remained a mystery. This result is contradictory to the general perception about the dropwise and filmwise modes of condensation under pure vapor conditions, but a plausible reason has not yet been addressed. Furthermore, because these investigations are limited to a narrow range of ambient conditions, making a firm conclusion about the most effective mode of condensation in humid air is difficult from Table \ref{tab1}.

To bridge this gap, we performed the experiments on two superhydrophobic and two superhydrophilic surfaces under humid air conditions in a much more comprehensive range of degrees of subcooling (10 - 26 $^\circ C$) and humidity ratio difference (5 - 45 \textit{g/kg of dry air}). Wherein, the humidity ratio difference ($\Delta \omega$) is defined as the difference between the specific humidity of the bulk humid air and the saturated vapor at the substrate temperature conditions. We observed that, depending on the specific combination of the experimental conditions investigated, filmwise mode on superhydrophilic surfaces can yield 57-333 \%  greater condensate yield than the dropwise mode on superhydrophobic surfaces. To substantiate these observations, a detailed assessment has been performed using basic condensation heat transfer theory for computing the thermal resistance offered by coating layer, condensate, liquid-vapor interface, diffusion layer, etc. \cite{le1966theory, kim2011}. This shows that in the presence of NCGs (specifically above 95 \%), the thermal resistance across the condensate and coating layer becomes negligible, and vapor diffusion becomes the dominant mechanism. We also investigated the different rivulet structures of the condensate forming on superhydrophilic surfaces due to the surface morphology, spreading dynamics, roughness, etc., and provide first insights into how they influence the condensation rate in a humid air environment. Surprisingly enough, the superhydrophilic surface with vertical rivulet channels outperformed those with branched rivulet channels by 10-20 \%. The current study also sheds light on the previously unknown benefits of the filmwise mode in humid air, which have far-reaching implications for developing efficient engineered interfaces for humidity harvesting. 

\section*{Results}\label{sec2}

\subsection*{Surface morphology}
\begin{figure*}[ht!]
    \centering
    \includegraphics[width=0.9\linewidth]{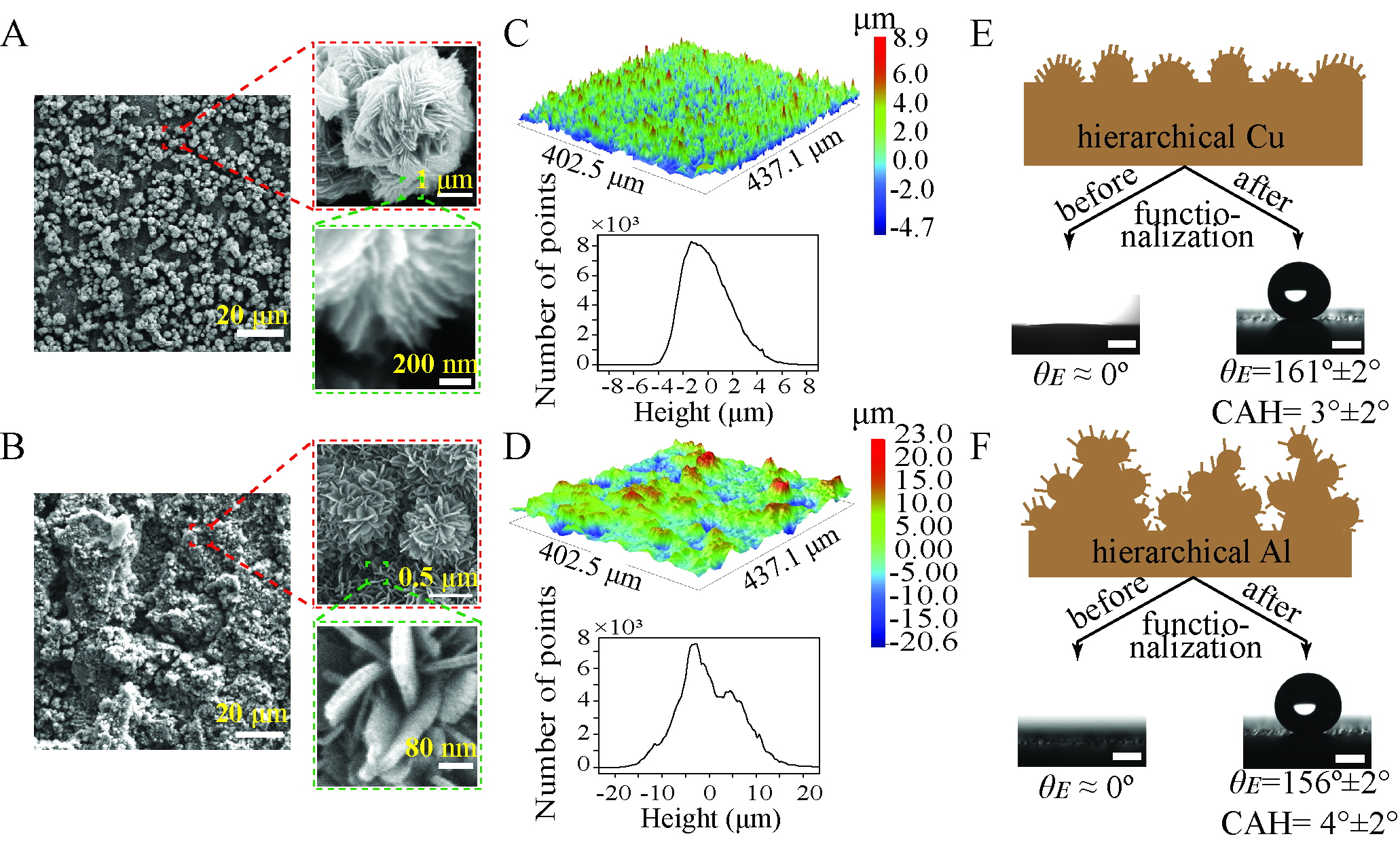}
    \caption{\textbf{Surface topography and characterization.} SEM micrograph of micro/nano textured (A) copper and (B) aluminum substrate at various resolutions. Profilometer micrograph of (C) copper and (D) aluminum substrates. The frequency distribution graph depicts the number of peaks and valleys at each bin size. The schematic illustration of the randomly generated micro/nano textures on (E) copper and (F) aluminum substrates. The scale bars in contact angle images are 1 $mm$.}    
    \label{fig:sem}
\end{figure*}

To investigate the condensation performance on completely wetting and non-wetting surfaces under humid air conditions, hierarchical surfaces of copper and aluminum were fabricated. Figures \ref{fig:sem}(A and B) show the surface morphology of the micro/nano textured copper and aluminum substrates, both of which exhibit superhydrophilic nature. The hierarchically rough substrates were functionalized with perfluorooctyl-triethoxy silane (PFOTES) to achieve superhydrophobicity. The surface morphology of the copper and aluminum superhydrophobic surfaces was similar to the corresponding superhydrophilic surfaces since the self-assembled monolayer of PFOTES does not modify the hierarchical structures \cite{yang2006dropwise}. The copper substrate consists of micro-flower shaped cupric oxide structures ranging in size from 2 - 4 $\mu m$, that are orderly arranged on the surface to some extent (Fig. \ref{fig:sem}A) \cite{zhang2003single,zhang2017enhanced}. The orders of the structures can be visualised in the profilometer micrograph, as shown in Fig. \ref{fig:sem}C. The spacing between the micro-flowers was in the range of 0.5 - 15 $\mu m$. Conversely, the surface features of the aluminum sample were random and consisted of micro-bumps of different sizes, as shown in Figs. \ref{fig:sem}(B and D). The height and width of these micro-bumps were in the range of 5 - 30 $\mu m$ and 5 - 60 $\mu m$, respectively. Flower-shaped features of size 0.5 - 1 $\mu m$ are found to surround the micro-bumps (Fig. \ref{fig:sem}B). The micro-flower structures on the copper substrate consist of thin sheets of nanostructures, and the aluminum substrate consists of blade-shaped nanostructures. 

The actual surface topography of the copper and aluminum substrates can be schematically represented based on the SEM and profilometer micrographs in Figs. \ref{fig:sem}(E and F), respectively. The average roughness of the modified copper and aluminum substrates was $R_a$ = 1.5 $\mu m$, $R_a$ = 5.3 $\mu m$, respectively. The frequency distributions in Figs. \ref{fig:sem}(C and D) show that $\sim99$ \% of the peaks and valleys lie in the range of $\pm4$ $\mu m$ and $\pm15$ $\mu m$ for copper and aluminum substrates, respectively. These  indicate that the depth of the textures is higher on the aluminium substrate than on the copper substrate. Prior to PFOTES functionalization, the contact angle of the aluminium and copper substrates was nearly zero, and a water drop dispensed on the substrate completely spread on the surface through hemiwicking. The equilibrium contact angle of the PFOTES-coated copper surface was $\theta$ = 161$\pm$ 2$^\circ$ and the contact angle hysteresis (CAH) was 3 $\pm$ 2 $^\circ$. The equilibrium contact angle of the superhydrophobic aluminum substrate was $\theta$ = 156 $\pm$ 2$^\circ$ and CAH=4 $\pm$ 2$^\circ$.

\subsection*{Condensation performance}
\begin{figure*}[ht!]
    \centering
    \includegraphics[width=0.9\linewidth]{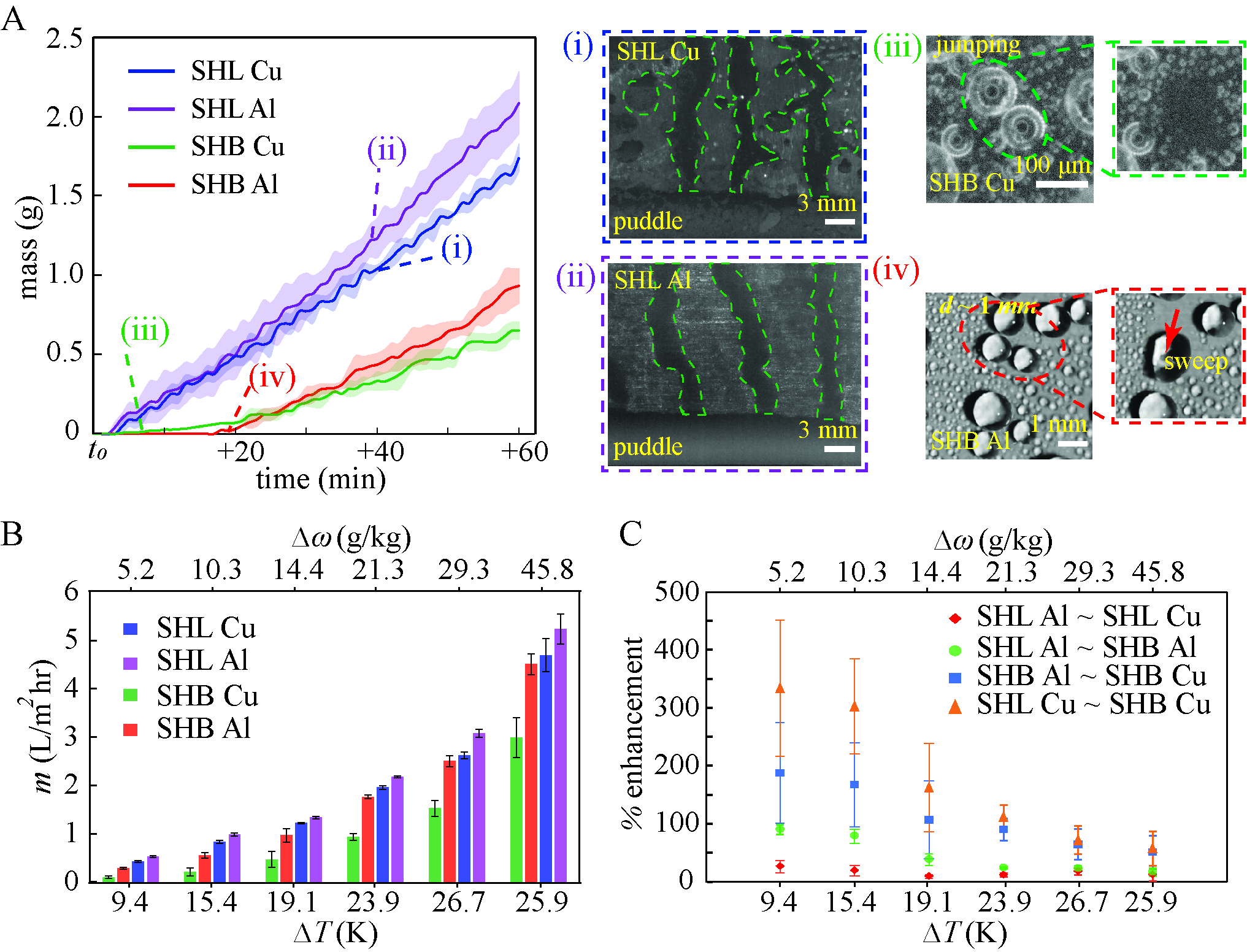}
    \caption{\textbf{Water yield from different substrates.} (A) The amount of water collected from the surfaces over time at a temperature of $T_{env}=30$ $^\circ C$, RH = 60 \%, and a surface temperature of $T_s$= 10 $\pm$ 0.5 $^\circ C$. The measurement was recorded at a time interval of 1 $minute$ using a microbalance of 10 $mg$ accuracy. $t_0$ is the time at which the surface temperature reaches the substrate temperature of $T_s$. The shaded area denotes the uncertainty in the average mass of water collected from three repeated experiments. Inset images show the filmwise mode of condensation on (i) copper superhydrophilic (SHL) (ii) aluminum superhydrophilic surface, (iii) droplet jumping on superhydrophobic (SHB) copper surface, and (iv) condensate removal from superhydrophobic aluminum surface. (B) The overall water collection rate of different surfaces under various environmental conditions is summarized in Table \ref{tab2}. (C) Percentage enhancement in condensation mass transfer under different environmental conditions for the surfaces between SHL Al $\sim$ SHL Cu, SHL Al $\sim$ SHB Al, SHB Al $\sim$ SHB Cu, and SHL Cu $\sim$ SHB Cu.}     
    \label{fig:mass}
\end{figure*}

Condensation experiments in the quiescent air environment were performed to compare the water collection rate from the fabricated surfaces at a degree of subcooling of 10 - 26 $^\circ C$ and a humidity ratio difference of 5 - 45 $g/kg$ \textit{of dry air}. The different experimental conditions chosen for this study are shown in Table \ref{tab2}. The condensation rate of the surfaces was evaluated by measuring the mass of the drained condensate from the surface using a microbalance at every experimental condition. Figure \ref{fig:mass}A depicts the amount of water collected in a container over time for different surfaces considered in this study. These experiments were performed at a temperature of $T_{env}=30$ $^\circ C$, RH = 60 \%, and a surface temperature of $T_s$= 10 $\pm$ 0.5 $^\circ C$. On superhydrophilic surfaces, the vapor nucleated easily due to the low energy barrier, and the condensate spread due to hemiwicking \cite{kim2016dynamics} and formed a thin film that drains through the vertical surface due to gravity (Fig. \ref{fig:mass}A (i) and (ii)). The drained condensate accumulated at the bottom of the surface as a puddle due to the pinning resistance at the bottom edge of the substrate. Therefrom, the condensate liquid was removed as the gravity force on the accumulated liquid eventually exceeded the pinning resistance. The first event of condensate drainage from superhydrophilic copper and aluminum substrates occurred almost the same time $\sim t_0+3$ $minutes$ for the size of the condensate plate considered here. The trend of the water collection was linear with time for both superhydrophilic copper and aluminum surfaces, but the rate of condensation of the superhydrophilic aluminum surface was found to be higher than on the superhydrophilic copper surface (see the plot in Fig. \ref{fig:mass}A).
\begin{table}[ht]
\centering
\resizebox{0.8\textwidth}{!}{
\begin{tabular}{cccccc}
\toprule
\begin{tabular}[c]{@{}c@{}}Environmental \\ temperature \\ $T_{env}$, $^\circ C$\end{tabular} & \begin{tabular}[c]{@{}c@{}}Relative \\ humidity \\ $\phi$, \%\end{tabular} & \begin{tabular}[c]{@{}c@{}}Surface \\ temperature \\ $T_{s}$, $^\circ C$\end{tabular} & \begin{tabular}[c]{@{}c@{}}Dew point \\ temperature \\ $T_{dew}$, $^\circ C$\end{tabular} & \begin{tabular}[c]{@{}c@{}}Degree of \\ subcooling \\ $\Delta T$, $^\circ C$\end{tabular} & \begin{tabular}[c]{@{}c@{}} Humidity \\ ratio \\ difference \\$\Delta \omega$, $g/kg$ \end{tabular}   \\ \midrule
20                                                                   & 75                                                           & 6 $\pm$ 0.5                                                              & 15.43                                                           & 9.43       & 5.17                                                        \\
30                                                                   & 60                                                           & 6 $\pm$ 0.5                                                              & 21.39                                                             & 15.39       & 10.26                                                        \\
30                                                                   & 75                                                           & 6 $\pm$ 0.5                                                              & 25.08                                                             & 19.08       & 14.39                                                        \\
35                                                                   & 75                                                           & 6 $\pm$ 0.5                                                              & 29.89               & 23.89       & 21.25                                                        \\
40                                                                   & 75                                                           & 8 $\pm$ 0.5                                                              & 34.71                   & 26.71       & 29.31                                                        \\
45                                                                   & 90                                                           & 17 $\pm$ 1                                                              & 42.97               & 25.97       & 45.82                                                        \\ \bottomrule
\end{tabular}}

\caption{\textbf{The experimental conditions considered in the present study using environmental chamber.} The chosen environmental conditions cover a wide range of degrees of subcooling and differences in humidity ratios that can be encountered with atmospheric water harvesting systems.}
\label{tab2}
\end{table}

Coalescence-induced droplet jumping was observed on superhydrophobic copper substrate \cite{boreyko2009self, miljkovic2013jumping} which predominantly caused the droplet removal (Fig. \ref{fig:mass}A(iii)). The coalescence-induced droplet jumping events have been identified as two-drop (Fig. S1a) or multi-drop jumping (Fig. S1b). Furthermore, droplet sweeping events were also observed sporadically on it (Fig. S2a). On the other hand, condensate drops from the superhydrophobic aluminum substrate were found to be removed by gravity-assisted sweeping mode (Fig. \ref{fig:mass}A(iv)). The jumping microdroplets can travel a longer distance from the substrate during their flight. For the purpose of measurement of condensation rate, it is important to ensure that the condensate from all the surfaces reaches the condensate collection system without any mass loss inflicted by jumping droplets falling outside the collector. The maximum possible velocity of the jumping droplet is in the order of $\sim30$ $cm/sec$ \cite{baba2020dropwise}. In the current experiments, the height between the top of the surface to the microbalance was around $\sim15$ $cm$. The horizontal distance traveled by the jumped droplet from the surface can be theoretically computed as $U\sqrt{2h/g}$ from the equations of motion, and the corresponding value is $\sim$5 $cm$. We used a petridish of diameter 9 $cm$ to collect the condensate from the substrate. Since the size of the petridish is larger than the maximum possible horizontal distance a jumping droplet can travel, we can safely assume that all the condensed droplets are collected, and there are no mass losses associated with jumping microdroplets. The influence of surface morphology on the microscale behavior of condensation on superhydrophobic copper and aluminum is detailed in the Supplementary Section S1. Coalescence of drops promptly ejected the condensate microdroplets from the superhydrophobic copper surface even before $t_0$ $minutes$. The mass of such microdroplets was difficult to measure experimentally since they weighed less than the microbalance resolution (10 $mg$). However, with the accumulation of multiple such droplets in the collection container, measurement was possible from $\sim t_0+4$ $minutes$ onwards, as shown in Fig. \ref{fig:mass}A. Interestingly, the onset of water collection on the superhydrophobic aluminum substrate started at $t_0+19$ $minutes$ despite the surface having a very low contact angle hysteresis (below 5$^\circ$). On the superhydrophobic aluminum surface, condensate was removed by gravity-assisted sweeping, and the drops exhibited a departure diameter of $\sim$1.4 $\pm$ 0.5 $mm$ above which they were drained by gravity from the surface (Fig. S2b). Although the drainage of condensate occurs on the superhydrophobic copper surface earlier than the superhydrophobic aluminum surface (Fig. \ref{fig:mass}A), the water collection from the superhydrophobic aluminum surface surpasses that from the superhydrophobic copper substrate beyond $t_0+27$ $minutes$. The water collection rate after the initial droplet drainage on the superhydrophobic aluminum surface is found to be higher than that on the superhydrophobic copper substrate because on the former the drainage of a single droplet causes additional removal of multiple droplets while sweeping.

A stable dropwise condensation mode was observed on superhydrophobic copper and aluminum substrates under all experimental conditions for a prolonged time, as shown in Fig. S4. In all cases, the superhydrophilic aluminum substrate had a higher water collection rate than all other surfaces, as shown in Fig. \ref{fig:mass}B. The water collection rate increases with the humidity ratio difference and degree of subcooling \cite{das2021filmwise}. The condensation rate of the superhydrophobic aluminum surface was found to be higher than that of the superhydrophobic copper surface at all the experimental conditions. The coalescence-induced droplet jumping mechanism played a dominant role in removal of the condensate from the superhydrophobic copper surface, while for the superhydrophobic aluminum surface, the drainage mechanism was primarily the sweeping. This suggests that the vertically oriented  surfaces exhibiting droplet jumping are less efficient for atmospheric humidity harvesting applications \cite{thomas2021condensation}. Interestingly enough, the same phenomenon has been strongly attributed to enhancing the condensation heat transfer in pure vapor conditions \cite{miljkovic2013jumping, wen2018three, liu2021coalescence}. During pure vapor condensation, the nano-engineered superhydrophobic surface led to flooding at higher subcooling due to an uncontrolled nucleation rate \cite{miljkovic2013jumping,wang2020density, olceroglu2016self}. However, the event of flooding was not observed on the superhydrophobic surfaces during the humid air condensation experiments at higher subcooling conditions even after 6 hours. This indicates that the presence of NCG hampers the nucleation of fresh embryos on the subcooled substrate and helps in the prevention of flooding on superhydrophobic surfaces at higher subcooling conditions. Figure \ref{fig:mass}C compares the percentage enhancement of condensation mass transfer between different surfaces under a wide range of environmental conditions. The superhydrophobic copper surface yielded 50 - 190 \% less condensate collection than the superhydrophobic aluminum surface since the surfaces exhibiting droplet jumping are less effective than the surfaces exhibiting sweeping mode of drainage in the atmospheric humid air environment \cite{thomas2021condensation}. The filmwise mode on the superhydrophilic aluminum surface yielded 16 - 90 \% more condensate collection than the dropwise mode on the superhydrophobic aluminum surface with gravity-assisted drainage. The dropwise condensation on the superhydrophobic copper substrate yielded 57 - 333 \% lower condensation rate than the superhydrophilic copper surface. A reduction in condensation rate by 333 \%  on superhydrophobic copper surface at low vapor content ($\Delta \omega=5.2$ $g/kg$) indicates that the superhydrophobic surfaces, which can show droplet jumping during condensation, are the least effective surfaces for humidity harvesting. 
\subsection*{Condensation on superhydrophilic surfaces: Role of rivulets}
Another important observation from this study is that the condensation rate is higher on the superhydrophilic aluminium surface than the superhydrophilic copper surface, although both surfaces had a contact angle of almost 0$^\circ$. The condensation rate of the superhydrophilic aluminum surface has shown an enhancement of 9 - 26 \% over the superhydrophilic copper surface. Importantly, at low vapor content ($\Delta \omega=$5.2 $g/kg$), the superhydrophilic aluminum surface showed an enhancement of nearly 26 \%, which is essential for the humidity harvesting system because getting a higher condensation rate at a low humidity ratio situation is critical \cite{gerasopoulos2018effects}. The influence of surface textures of superhydrophobic surfaces on condensation was extensively investigated in the past \cite{sharma2017growth, chen2015exploiting, rykaczewski2013multimode, olceroglu2016self, orejon2019dropwise, winter2020dewetting, zhao2015condensate, sharma2019self} because they exhibited enhanced performance under pure vapor conditions. However, such influences of superhydrophilic or hydrophilic surfaces were not reported. These results imply that the significance of the micro/nano features on the superhydrophilic surface has an impact on condensation performance in the humid air.    
    
\begin{figure*}[ht!]
    \centering
    \includegraphics[width=0.9\linewidth]{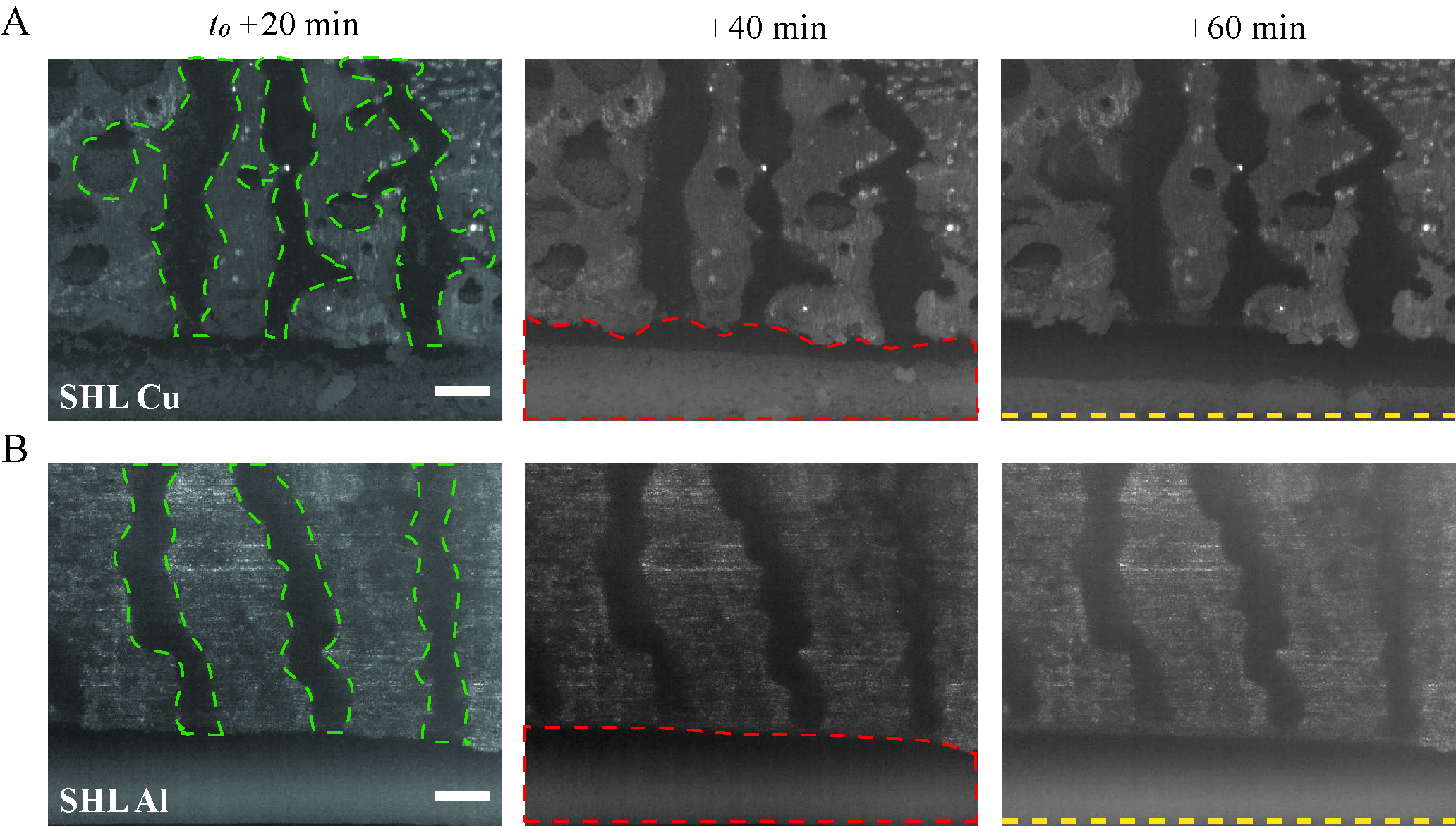}
    \caption{\textbf{Rivulets on superhydrophilic surfaces.} Time-lapse snapshots of condensation on superhydrophilic (SHL) (A) copper and (B) aluminum substrates at $T_{env}$= 27 $^\circ C$, $RH$= 60 \%, $T_{dew}$= 18.6 $^\circ C$ and $T_s$= 6 $\pm$ 0.5 $^\circ C$. $t_0$ is the time at which the surface temperature reaches the set value. Here, the green and red dashed regions represent the rivulet and puddle regimes, respectively. The Yellow dashed line represents the bottom edge of the substrate. For clear visualization of different regimes on superhydrophilic surfaces during condensation, we used IR imaging (FLIR-A655sc) to produce the above image. The scale bar is 3 $mm$.}    
    \label{fig:shl}
\end{figure*}

At the initial stage of condensation on superhydrophilic vertical surfaces, the nucleated drops spread over the surface and form the rivulet regime in which condensate is found to be drained downward by gravity. The condensate, drained through rivulets, accumulates at the bottom of the substrate as a puddle (marked as a red dashed line in Fig. \ref{fig:shl}) due to the pinning force offered by the bottom edge. The puddle volume grows over time, and it eventually drains off the surface when gravity overcomes the pinning force at the bottom-edge. The size of the puddle reduces right after the drainage and then again increases with time until the next shedding happens. The nucleated condensate observed in the thin-film regime (the region outside of the puddle and rivulet regimes) spreads and merges with the rivulet or puddle. Gravity effects are negligible in the thin-film region. The rivulet regime has a thicker condensate film than the thin film regime. The shape of the rivulet depends on the surface topography and heterogeneity. The observed rivulet shape differed between copper and aluminum superhydrophilic substrates, and did not vary with time once they reached a steady-state (Fig. \ref{fig:shl}). The rivulets on the superhydrophilic aluminum substrate are mostly vertical channels, whereas the rivulets on the superhydrophilic copper surface consist of multiple branches that are inclined or horizontal and connected to the vertical channel. The condensate present in the vertical rivulet channel is smoothly drained due to gravity. The condensate accumulated in the rivulet branches is either stagnant or flowing with a lower velocity towards the central vertical channel due to the flow resistance offered from the accumulated condensate in the central channel. Thus, the overall drainage rate on the superhydrophilic copper substrate is reduced.

The above observations clearly elucidate that condensation performance of an superhydrophilic surface under humid air condition is significantly influenced by rivulet formation and the dynamics of the drainage mechanism through the rivulets. The spreading of condensate in the thin film region leads to the continuous transport of condensate from the thin film regime to the rivulet regime, which promotes homogeneous nucleation in the thin-film regime. As a result, at steady state, the rate of condensate formation in the thin-film regime is higher than that in the rivulet regime. Overall, it is clear that an optimal superhydrophilic surface with vertically shaped rivulet channels can improve the condensation rate for humidity harvesting.

Notably, under all experimental conditions considered in this work, condensation mass transfer in the filmwise mode of condensation on the superhydrophilic surface was greater than that in the dropwise mode of condensation on the superhydrophobic surface (Fig. \ref{fig:mass}B). This observation contradicts the general perception of condensation under pure steam conditions  \cite{schmidt1930condensation, le1965experimental, cha2020dropwise}. Although multiple research works summarized in Table \ref{tab1} showed observations similar to our foregoing experimental results, the reason for this observation was not explained in any of these studies. A detailed comparison between condensation phenomena in the pure vapor and NCG environments is necessary and presented next.

\subsection*{Influence of non-condensable gases}   
To understand the influence of NCG during condensation, a theoretical analysis of a single condensing drop \cite{le1966theory, kim2011} and a condensate film \cite{nusselt1916surface} was carried out for a wide range of NCG concentrations. This section compares the nucleation performance of different surfaces and the individual contribution of temperature drop due to different thermal resistances at a moderate and exemplar subcooling of 10 $K$. 

The nucleation rate of a substrate is defined as the probability of the liquid embryo that is continuing to grow on a nucleation site without evaporation during vapor-liquid phase conversion. The heterogeneous nucleation rate per unit area ($dN/dt$) can be expressed as \cite{beysens2022physics},  

\begin{equation}
    \frac{dN}{dt}=\frac{R_0}{3}Ae^{\frac{-W}{k_BT_s}}
\end{equation}

where $R_0$ is the radius of the water molecule, $R_0 \approx 1.375\times10^{-10}$ $m$, $A$ is the Arrhenius prefactor, $W$ is the total energy required for the formation of a liquid nuclei, $k_B$ is the Boltzmann constant and $T_{s}$ is the substrate temperature. 

\begin{figure*}[ht!]
    \centering
    \includegraphics[width=0.9\linewidth]{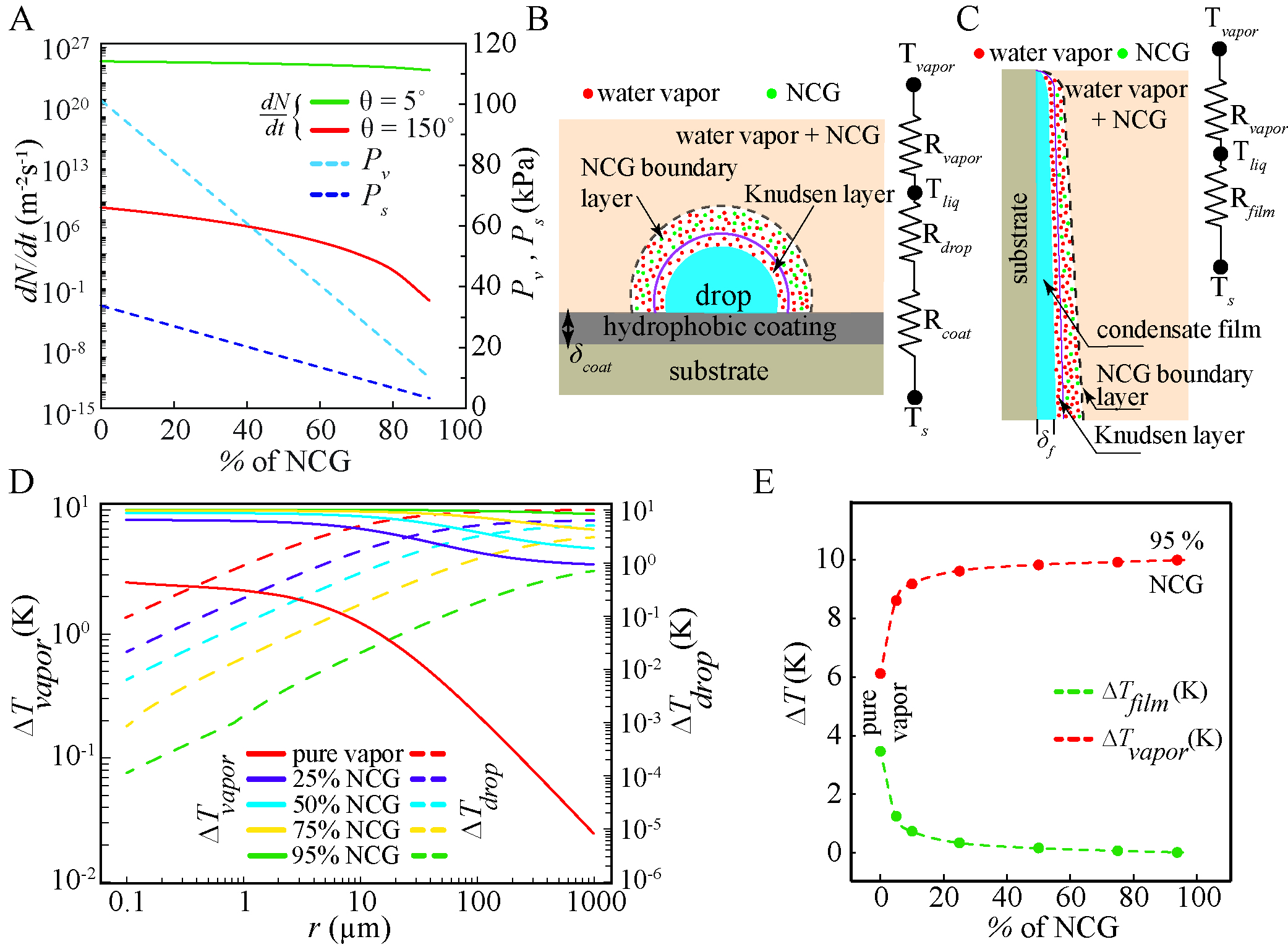}
    \caption{\textbf{Nucleation rate and temperature drops in different modes of condensation.} (A) The influence of NCG concentration and surface wettability on heterogeneous nucleation rate for a fixed supersaturation ratio, $S=P_v/P_s=2$. Schematic of heat transfer model with thermal resistance network across (B) a single droplet during dropwise condensation and (C) a film during filmwise condensation in the presence of NCG. The influence of NCG concentration on temperature drop across the liquid (film or droplet) due to conduction ($\Delta T_{drop/film}$) and temperature drop across the vapor region ($\Delta T_{vapor}$) during (D) dropwise mode, and (E) filmwise mode. The thermal conductivity ($k_{coat}$) and the thickness of the hydrophobic layer coating ($\delta_{coat}$) are approximated as 0.2 $W/mK$ and 1 $\mu m$ respectively \cite{kim2011}. $\delta_f$ is the average condensate film thickness and it is approximated as 25 $\mu m$ \cite{zhang2021analytical}. Refer Supplementary section S2 for more details.}    
    \label{fig:model}
\end{figure*}

The vapor pressure is significantly reduced with the increase in NCG concentration (Figure \ref{fig:model}A). It also reduces the heterogeneous nucleation rate \cite{beysens2022physics}. The accumulation of NCG near the subcooled surface can further reduce water vapor diffusion from the bulk region to the vapor-liquid interface, and thus, hinders the nucleation rate. Figure \ref{fig:model}A compares the influence of NCG concentration on the nucleation rate for extreme wettability (such as superhydrophilic and superhydrophobic) surfaces at a supersaturation ratio ($S$) of 2. Supersaturation ratio is defined as partial pressure of the water vapor ($P_v$) to the saturation pressure corresponding to the substrate temperature $T_s$ ($P_s$). As can be seen from Fig. \ref{fig:model}A, the nucleation rate decreases with NCG concentration; at 90 \% NCG concentration the nucleation rate of the superhydrophilic surface reaches a value that is one order of magnitude lower than that for the pure vapor condition. On the other hand, the nucleation rate on the superhydrophobic surface reaches a value that is eleven orders of magnitude lower than that of the pure vapor condition, as shown in Fig. \ref{fig:model}A. As a result, the nucleation occurs much slowly from humid air than from pure vapour on superhydrophobic surfaces. The interplay between the phase transition of vapor molecules and vapor diffusion from the bulk fluid makes the nucleation phenomena complex in the presence of NCG. The probability of activating a nucleation site in the NCG environment depends significantly on the localised vapor content and vapor diffusion rate across the NCG diffusion layer.  

Figures \ref{fig:model}(B and C) show a schematic representation of the thermal resistance network model in the presence of NCG across a single droplet for the dropwise condensation and across the film for the filmwise condensation. The schematics consist of three regions, namely hydrophobic coating region, the condensed drop/film region, and the vapor region. The vapor region consists of interfacial region called Knudsen layer, the diffusion layer and the bulk vapor-NCG mixture region. In the Knudsen layer, the kinetic theory of gases governs the transfer of vapor molecules to the liquid drop interface. The region outside the Knudsen layer is known as the diffusion layer, and during condensation, a concentration gradient of vapour exists in this region. The thickness of the diffusion layer depends both on the hydrodynamic parameter (the free stream velocity for a forced flow and the Grashof number in case of a thermo-gravitational flow), and the thermophysical properties (e.g., thermal and mass diffusivities). For the present cases, the hydrodynamic influence on the vapor-NCG boundary layer may be neglected since the pertinent thermo-gravitational Grashof number, $Gr=g\beta \Delta T L_c^3/\nu^2$ is small ($\sim10^{5}-10^{6}$), where $g$ is the acceleration due to gravity, $\beta$ is the co-efficient of thermal expansion, $L_c$ is the characteristic length scale, and $\nu$ is the kinematic viscosity. On the other hand, a change in the degree of subcooling and the bulk vapor concentration alters the average mass and thermal diffusivities, thereby altering the diffusion layer thickness. The thickness of the diffusion layer increases as the vapour concentration of the mixture decreases. Outside the Knudsen outer interface, combined mass diffusion and energy conservation laws govern the transport. For condensation under pure steam scenario, the vapour in the bulk region remains in an equilibrium state, and the interface temperature of the drop is always maintained at the saturation temperature corresponding to the vapour pressure. However, presence of NCGs adds heat and mass transfer resistances, thus rendering the actual prediction of interface temperature difficult. It can only be predicted by solving coupled mass and energy conservation equations \cite{luo2016new,zheng2018modeling,liu2020heat}. The total temperature difference ($\Delta T$) between the cold surface and the bulk vapor-NCG mixture is expressed as follows:
\begin{equation}
\Delta T= \Delta T_{coat}+\Delta T_{drop/film}+\Delta T_{vapor}
\label{eq:dT}
\end{equation}
where $\Delta T_{coat}$, $\Delta T_{drop/film}$, $\Delta T_{vapor}$ are the temperature drops across the promoter layer coating, liquid drop/film due to conduction, vapor region respectively. It is important to note that the $\Delta T_{vapor}$ includes the temperature drop due to curvature resistance ($\Delta T_{c}$, see Eq. S11) and interfacial temperature drop ($\Delta T_{int}$, see Eq. S12). The estimation of each temperature drop during dropwise and filmwise mode of condensation is calculated from the numerical iteration of analytical expressions or an existing correlation (Supplementary Section S2). 

The amount of temperature drop over the condensate drop and vapor region for a drop of radius 100 $nm$ - 1 $mm$ during dropwise mode condensation of pure vapour and the vapor-NCG mixture is shown in Figure \ref{fig:model}D. In the dropwise mode of condensation, the temperature drop across the vapor region is significantly higher for the droplets of radius below 0.1 $\mu m$ irrespective of the NCG concentration. This is because the curvature component ($\Delta T_c$) offers greater thermal resistance for smaller droplets. The contribution of these thermal resistances to the overall temperature drop decreases as drop size increases. In the case of pure vapor condensation, the temperature drop across the droplet is significant except for the case of smaller droplets of radius below 0.1 $\mu m$ and it reaches above 90 \% of the total subcooling for droplets with a radius greater than 200 $\mu m$. With a rise in NCG concentration, the temperature drop across the drop due to conduction decreases. With  95 \% NCG (for example, atmospheric humid air), the temperature drop within the condensate droplet is entirely negligible for any droplet size (Fig. \ref{fig:model}D), and the magnitude of temperature drop across the coating is negligible (Fig. S3), even at a coating thickness of 10 $\mu m$. For filmwise condensation, variation in the temperature drop across the film and vapor regions with different NCG concentrations is shown in Fig. \ref{fig:model}E. Both the temperature drops (across the film and vapor region) are significant, and they occur in the same order as for pure vapour condensation. However, the temperature drop across the film decreases with NCG concentration and becomes negligible at concentration greater than 50 \%. Moreover, the drop occurs primarily at the vapor region. 

Therefore, regardless of the mode of condensation, the heat transfer performance during humid air condensation is primarily controlled by the rate of diffusion of the vapour molecules through the NCG diffusion layer, and the thermal resistance offered by the condensate drop or the coating is insignificant. The heat transfer performance during the filmwise mode is poor in a pure vapor environment due to the significant thermal resistance across the condensate film. In humid air condensation however, a similar conclusion is incorrect because the thermal resistance across the film/drop is negligible and the interface temperature of the film/drop is close to the substrate temperature \cite{zheng2018modeling}. In addition to this, the nucleation rate of the superhydrophilic surface is twenty seven orders of magnitude higher than the superhydrophobic surface (Fig. \ref{fig:model}A). As a result, rejuvenation of heterogeneous nucleation sites after the drop removal is energetically expensive for superhydrophobic surfaces in humid air conditions. Based on these understandings and observations from the experiments, we can interpret that superhydrophilic surfaces can show better condensation heat transfer performance over the superhydrophobic surface in the humid air environment with more than 95 \% NCG, and they can outperform in atmospheric humidity harvesting applications.

\section*{Discussion}\label{sec3}

The major challenge in humidity harvesting devices is obtaining the required amount of water yield from the atmospheric humid air on a daily basis. The air temperature and relative humidity vary throughout the day, and the water yield from the surface can be drastically reduced, particularly when the humidity ratio is low. As a result, increasing water yield under low humidity conditions is critical for obtaining the desired water yield from the humidity harvesting system. The superior condensation rates on superhydrophilic surfaces compared to the superhydrophobic surface at low humidity ratios is appealing for humidity harvesting systems. From other aspects also, such as the cost associated with surface fabrication, scalability of the surface fabrication technique, and chemical contamination possibilities of condensate with hydrophobic coating material, superhydrophilic surface appears to be more suitable than the superhydrophobic surfaces for humidity harvesting.     
  
In summary, this study investigated the condensation performance of vertically oriented superhydrophilic and superhydrophobic surfaces in a wide range of humid air environments. Interestingly, at a vapor concentration of below 5 \%, we found that the filmwise mode of condensation on a superhydrophilic surface shows superior water collection than the dropwise mode of condensation observed on a superhydrophobic surface. Theoretical analyses indicate that under humid air conditions, the conduction thermal resistance across the condensate liquid or the surface coatings is negligible, and the condensation performance is driven by diffusion and drainage rates. Hence, film formation does not significantly affect the overall condensation performance unlike what usually happens in the pure steam environment \cite{schmidt1930condensation, le1965experimental}. Besides, the very high nucleation rate on a superhydrophilic surface can substantially increase the condensation rate. However a wettability-patterned design with optimum nucleation rate and drainage rate may produce higher water yield than homogeneous superhydrophilic surface. In effect, our study provides a new mechanistic understanding of why the filmwise mode of condensation is better for humidity harvesting, as well as give an insights into exploiting new physics of rivulets to maximise this effectiveness. The key findings of this study can usher in a new pathway to the development of efficiently engineered surfaces for humidity harvesting systems.

\section*{Methods}\label{sec4}

\subsection*{Surface preparation and characterization}
Aluminum alloy (Al-6061) and copper (99.9 \% purity) plates of size 80 $mm$ $\times$ 40 $mm$ $\times$ 2 $mm$ were used as the substrate for the condensation experiments. First, the substrates were mechanically polished with sand paper of grade 220 to 2000 sequentially. Then the polished substrate was washed with DI water before being ultrasonically cleaned for 10 minutes in a solution of DI water, ethanol, and acetone. The clean copper substrate was kept in a 3$M$ $HNO_3$ (Merck, Emplura grade) solution for 10 minutes to remove the native oxides, and further rinsed with DI water and dried with nitrogen gas. Thereafter, the substrate was oxidised in an aqueous solution of 2$M$ $NaOH$ (Sigma Aldrich, ACS reagent grade pellets) and 0.1$M$ $K_2S_2O_8$ (Sigma Aldrich, ACS reagent grade) for 1 hour \cite{chen2015exploiting}, yielding a superhydrophilic surfaces. Next, the substrate was rinsed with DI water, and dried with nitrogen gas, and kept in the hot air oven for 30 minutes at a temperature of 150 $^\circ C$. To achieve superhydrophobicity, the oxidised copper substrate was dip-coated for 4 hours in an ethanol solution with 0.5 \% perfluorooctyl-triethoxy silane (Sigma Aldrich) and dried in the hot air oven at 150 $^\circ C$ for 1 hour. 

The aluminium substrate was microtextured by dipping it in a 3$M$ solution of $HCl$ (Merck, Emplura grade) for 5 minutes and then rinsing it with DI water. The chemically etched aluminum substrate was further immersed in a hot DI water bath at a temperature of 100 $^\circ C$ for 30 minutes to generate the nanostructures above the microstructures \cite{vedder1969aluminum+}. The aluminum substrate became superhydrophilic after those fabrication steps. Next, the substrate was functionalized with 0.5 \% perfluorooctyl-triethoxy silane (PFOTES) in ethanol solution for 4 hours and then dried in the laboratory environment for 12 hours for achieving superhydrophobicity. 

A high-resolution scanning electron microscope (Inspect F-FEI) was used to examine the surface morphology. The roughness of the modified substrate was measured using a non-contact type profilometer (NT1100-Wyco). The static contact angle of water on the substrate was measured with a goniometer (Holmarc) by placing a sessile drop of volume 5 $\mu L$ using a micro-pipette. The dynamic contact angles were determined by injecting/drawing DI water into/from the sessile drop with a syringe pump at a creeping flow rate of 0.1 $\mu L/sec$ \cite{peng2018all}. Each reported contact angle comprises of the average of six measurements taken at various locations on the substrate.
           
\subsection*{Experimental procedures}
The condensation experiments for a vertically oriented surface were performed in an environmental chamber (PR2J-Espec) of size 500 $\times$ 750 $\times$ 600 $mm^3$ at different environmental conditions as shown in Fig. S5. The condensing substrate was placed at the centre of the environmental chamber to avoid the boundary wall effects. The fluctuations in air temperature and relative humidity were $\pm0.5$ $^\circ C$ and $\pm3$ \%, respectively. The substrates under test were attached to a Peltier (CP061HT-Tetech) element with thermal conductive tape and the remaining area of the Peltier element was covered with nitrile foam to avoid condensation on the Peltier cooling plate. The temperature of the Peltier element was controlled using a PID controller. Four holes of size 0.8 $mm$ were drilled on the lateral side of the substrate for the insertion of K-type thermocouples (Omega, 0.13 $mm$ bead diameter) for surface temperature measurement. The temperature from the thermocouples was logged onto a data acquisition system (DAQ970A-Keysight) at every minute. The condensation experiments were performed for 2-3 hours, and the condensed water was collected in a Petri dish with a diameter of 9 $cm$. This ensured that the condensate from all the surfaces reached the container while jumping or sweeping. The mass of the water from the Petri dish was measured in a microbalance (SPX622-Ohaus) with a precision of 10 $mg$. At least three experiments were performed to calculate the average condensation rate on a surface. The condensation images were captured with a DSLR camera (D750-Nikon) and a 105 $mm$ macro lens at 30 $fps$. An LED spotlight was used to illuminate the imaging area (CW8-Moritex).

\section*{Data Availability}
The data that support the findings of this study are available from the corresponding authors upon request.
\section*{Author Contributions}
T.M.T and P.S.M conceived the initial idea of the research. T.M.T performed the fabrication and characterization of the materials, designed the experimental setup, performed the experiments, analyzed the results, and wrote the Matlab code for the numerical model. R.G conceived the idea of numerical modelling. All authors contributed in writing the manuscript. P.S.M, R.G and M.K.T guided this research.
\section*{Acknowledgements}
P.S.M acknowledges the Science and Engineering Research Board (SERB) of Government of India (Project number ECR/2018/001806) for funding the work. P.S.M also acknowledges the partial funding through the Indian Institute of Technology Madras to the Micro Nano Bio Fluidics Group for Institutions of Eminence scheme of Ministry of Education, Government of India (Sanction. No: $11/9/2019-U.3(A)$). M.K.T. also acknowledges support by the Wellcome/EPSRC Centre for Interventional and Surgical Sciences (WEISS) (203145Z/16/Z), Wolfson Foundation and Royal Society for his Royal Society Wolfson Fellowship, and the NICEDROPS project supported by the European Research Council (ERC) under the European Union's Horizon 2020 research and innovation programme under grant agreement No. 714712. Authors also acknowledge Dr. Arvind Pattamatta, IIT Madras for providing the access of IR camera facility.

\medskip

\bibliographystyle{unsrt}
\bibliography{manuscript}

\begin{thebibliography}{10}

\bibitem{mekonnen2016four}
Mesfin~M Mekonnen and Arjen~Y Hoekstra.
\newblock Four billion people facing severe water scarcity.
\newblock {\em Science Advances}, 2(2):e1500323, 2016.

\bibitem{tu2018progress}
Yaodong Tu, Ruzhu Wang, Yannan Zhang, and Jiayun Wang.
\newblock Progress and expectation of atmospheric water harvesting.
\newblock {\em Joule}, 2(8):1452--1475, 2018.

\bibitem{nagar2020clean}
Ankit Nagar and Thalappil Pradeep.
\newblock Clean water through nanotechnology: Needs, gaps, and fulfillment.
\newblock {\em ACS Nano}, 14(6):6420--6435, 2020.

\bibitem{haechler2021exploiting}
Iwan Haechler, Hyunchul Park, Gabriel Schnoering, Tobias Gulich, Mathieu
  Rohner, Abinash Tripathy, Athanasios Milionis, Thomas~M Schutzius, and Dimos
  Poulikakos.
\newblock Exploiting radiative cooling for uninterrupted 24-hour water
  harvesting from the atmosphere.
\newblock {\em Science Advances}, 7(26):eabf3978, 2021.

\bibitem{kroger1968condensation}
Detlev~G Kroger and Warren~M Rohsenow.
\newblock Condensation heat transfer in the presence of a non-condensable gas.
\newblock {\em International Journal of Heat and Mass Transfer}, 11(1):15--26,
  1968.

\bibitem{zheng2018modeling}
Shaofei Zheng, Ferdinand Eimann, Christian Philipp, Tobias Fieback, and Ulrich
  Gross.
\newblock Modeling of heat and mass transfer for dropwise condensation of moist
  air and the experimental validation.
\newblock {\em International Journal of Heat and Mass Transfer}, 120:879--894,
  2018.

\bibitem{othmer1929condensation}
Donald~Frederick Othmer.
\newblock The condensation of steam.
\newblock {\em Industrial \& Engineering Chemistry}, 21(6):576--583, 1929.

\bibitem{slegers1970laminar}
L~Slegers and RA~Seban.
\newblock Laminar film condensation of steam containing small concentrations of
  air.
\newblock {\em International Journal of Heat and Mass Transfer},
  13(12):1941--1947, 1970.

\bibitem{oestreich2019experimental}
JL~Oestreich, CWM Van Der~Geld, JL~Goes Oliveira, and AK~Da~Silva.
\newblock Experimental condensation study of vertical superhydrophobic surfaces
  assisted by hydrophilic constructal-like patterns.
\newblock {\em International Journal of Thermal Sciences}, 135:319--330, 2019.

\bibitem{thomas2021condensation}
Tibin~M Thomas and Pallab Sinha~Mahapatra.
\newblock Condensation of humid air on superhydrophobic surfaces: Effect of
  nanocoatings on a hierarchical interface.
\newblock {\em Langmuir}, 37(44):12767--12780, 2021.

\bibitem{schmidt1930condensation}
E~Schmidt, W~Schurig, and W~Sellschopp.
\newblock Condensation of water vapour in film-and drop form.
\newblock {\em Technische Mechanik und Thermodynamik}, 1:53--63, 1930.

\bibitem{le1965experimental}
EJ~Le~Fevre and JW~Rose.
\newblock An experimental study of heat transfer by dropwise condensation.
\newblock {\em International Journal of Heat and Mass Transfer},
  8(8):1117--1133, 1965.

\bibitem{cha2020dropwise}
Hyeongyun Cha, Hamed Vahabi, Alex Wu, Shreyas Chavan, Moon-Kyung Kim, Soumyadip
  Sett, Stephen~A Bosch, Wei Wang, Arun~K Kota, and Nenad Miljkovic.
\newblock Dropwise condensation on solid hydrophilic surfaces.
\newblock {\em Science Advances}, 6(2):eaax0746, 2020.

\bibitem{choo2015water}
Soyoung Choo, Hak-Jong Choi, and Heon Lee.
\newblock Water-collecting behavior of nanostructured surfaces with special
  wettability.
\newblock {\em Applied Surface Science}, 324:563--568, 2015.

\bibitem{mahapatra2016key}
Pallab~Sinha Mahapatra, Aritra Ghosh, Ranjan Ganguly, and Constantine~M
  Megaridis.
\newblock Key design and operating parameters for enhancing dropwise
  condensation through wettability patterning.
\newblock {\em International Journal of Heat and Mass Transfer}, 92:877--883,
  2016.

\bibitem{seo2016effects}
Donghyun Seo, Junghun Lee, Choongyeop Lee, and Youngsuk Nam.
\newblock The effects of surface wettability on the fog and dew moisture
  harvesting performance on tubular surfaces.
\newblock {\em Scientific Reports}, 6(1):1--11, 2016.

\bibitem{hou2020tunable}
Kongyang Hou, Xiaoyang Li, Qiang Li, and Xuemei Chen.
\newblock Tunable wetting patterns on superhydrophilic/superhydrophobic hybrid
  surfaces for enhanced dew-harvesting efficacy.
\newblock {\em Advanced Materials Interfaces}, 7(2):1901683, 2020.

\bibitem{pinheiro2019water}
RA~Pinheiro, DD~Damm, AA~Silva, RM~Volu, KF~Almeida, FM~Rosa, VJ~Trava-Airoldi,
  and EJ~Corat.
\newblock Water vapor condensation from atmospheric air by super-hydrophobic
  vacnts growth on stainless steel pipes.
\newblock {\em MRS Advances}, 4(35):1929--1936, 2019.

\bibitem{pinheiro2020vertically}
Romario~Araujo Pinheiro, Filipe~Menezes Rosa, Ren{\^e}~Martins Vol{\'u},
  Get{\'u}lio de~Vasconcelos, Vladimir~Jesus Trava-Airoldi, and Evaldo~Jos{\'e}
  Corat.
\newblock Vertically aligned carbon nanotubes (vacnt) surfaces coated with
  polyethylene for enhanced dew harvesting.
\newblock {\em Diamond and Related Materials}, 107:107837, 2020.

\bibitem{gupta2019background}
Rohit Gupta, Chayan Das, Amitava Datta, and Ranjan Ganguly.
\newblock Background oriented schlieren (bos) imaging of condensation from
  humid air on wettability-engineered surfaces.
\newblock {\em Experimental Thermal and Fluid Science}, 109:109859, 2019.

\bibitem{lee2020improved}
Junbeom Lee, Seunghwan Lee, and Jaeseon Lee.
\newblock Improved humid air condensation heat transfer through promoting
  condensate drainage on vertically stripe patterned bi-philic surfaces.
\newblock {\em International Journal of Heat and Mass Transfer}, 160:120206,
  2020.

\bibitem{lee2012water}
Anna Lee, Myoung-Woon Moon, Hyuneui Lim, Wan-Doo Kim, and Ho-Young Kim.
\newblock Water harvest via dewing.
\newblock {\em Langmuir}, 28(27):10183--10191, 2012.

\bibitem{nagar2021scalable}
Ankit Nagar, Ramesh Kumar, Pillalamarri Srikrishnarka, Tiju Thomas, and
  Thalappil Pradeep.
\newblock Scalable drop-to-film condensation on a nanostructured hierarchical
  surface for enhanced humidity harvesting.
\newblock {\em ACS Applied Nano Materials}, 4(2):1540--1550, 2021.

\bibitem{sharma2017growth}
Chander~Shekhar Sharma, Juliette Combe, Markus Giger, Theo Emmerich, and Dimos
  Poulikakos.
\newblock Growth rates and spontaneous navigation of condensate droplets
  through randomly structured textures.
\newblock {\em ACS Nano}, 11(2):1673--1682, 2017.

\bibitem{chen2015exploiting}
Xuemei Chen, Justin~A Weibel, and Suresh~V Garimella.
\newblock Exploiting microscale roughness on hierarchical superhydrophobic
  copper surfaces for enhanced dropwise condensation.
\newblock {\em Advanced Materials Interfaces}, 2(3):1400480, 2015.

\bibitem{rykaczewski2013multimode}
Konrad Rykaczewski, Adam~T Paxson, Sushant Anand, Xuemei Chen, Zuankai Wang,
  and Kripa~K Varanasi.
\newblock Multimode multidrop serial coalescence effects during condensation on
  hierarchical superhydrophobic surfaces.
\newblock {\em Langmuir}, 29(3):881--891, 2013.

\bibitem{olceroglu2016self}
Emre Olceroglu and Matthew McCarthy.
\newblock Self-organization of microscale condensate for delayed flooding of
  nanostructured superhydrophobic surfaces.
\newblock {\em ACS Applied Materials \& Interfaces}, 8(8):5729--5736, 2016.

\bibitem{orejon2019dropwise}
Daniel Orejon, Alexandros Askounis, Yasuyuki Takata, and Daniel Attinger.
\newblock Dropwise condensation on multiscale bioinspired metallic surfaces
  with nanofeatures.
\newblock {\em ACS Applied Materials \& Interfaces}, 11(27):24735--24750, 2019.

\bibitem{winter2020dewetting}
Rebecca~L Winter and Matthew McCarthy.
\newblock Dewetting from amphiphilic minichannel surfaces during condensation.
\newblock {\em ACS Applied Materials \& Interfaces}, 12(6):7815--7825, 2020.

\bibitem{zhao2015condensate}
Ye~Zhao, Yuting Luo, Juan Li, Fei Yin, Jie Zhu, and Xuefeng Gao.
\newblock Condensate microdrop self-propelling aluminum surfaces based on
  controllable fabrication of alumina rod-capped nanopores.
\newblock {\em ACS Applied Materials \& Interfaces}, 7(21):11079--11082, 2015.

\bibitem{sharma2019self}
Chander~Shekhar Sharma, Cheuk Wing~Edmond Lam, Athanasios Milionis, Hadi
  Eghlidi, and Dimos Poulikakos.
\newblock Self-sustained cascading coalescence in surface condensation.
\newblock {\em ACS Applied Materials \& Interfaces}, 11(30):27435--27442, 2019.

\bibitem{le1966theory}
EJ~Le~Fevre and John~W Rose.
\newblock A theory of heat transfer by dropwise condensation.
\newblock In {\em International Heat Transfer Conference Digital Library}.
  Begel House Inc., 1966.

\bibitem{kim2011}
Sunwoo Kim and Kwang~J. Kim.
\newblock {Dropwise condensation modeling suitable for superhydrophobic
  surfaces}.
\newblock {\em Journal of Heat Transfer}, 133(8), 05 2011.

\bibitem{yang2006dropwise}
Qingfeng Yang and Anzhong Gu.
\newblock Dropwise condensation on sam and electroless composite coating
  surfaces.
\newblock {\em Journal of Chemical Engineering of Japan}, 39(8):826--830, 2006.

\bibitem{zhang2003single}
Weixin Zhang, Xiaogang Wen, Shihe Yang, Yolande Berta, and Z~Lin Wang.
\newblock Single-crystalline scroll-type nanotube arrays of copper hydroxide
  synthesized at room temperature.
\newblock {\em Advanced Materials}, 15(10):822--825, 2003.

\bibitem{zhang2017enhanced}
Peng Zhang, Yota Maeda, Fengyong Lv, Yasuyuki Takata, and Daniel Orejon.
\newblock Enhanced coalescence-induced droplet-jumping on nanostructured
  superhydrophobic surfaces in the absence of microstructures.
\newblock {\em ACS Applied Materials \& Interfaces}, 9(40):35391--35403, 2017.

\bibitem{kim2016dynamics}
Jungchul Kim, Myoung-Woon Moon, and Ho-Young Kim.
\newblock Dynamics of hemiwicking.
\newblock {\em Journal of Fluid Mechanics}, 800:57--71, 2016.

\bibitem{boreyko2009self}
Jonathan~B Boreyko and Chuan-Hua Chen.
\newblock Self-propelled dropwise condensate on superhydrophobic surfaces.
\newblock {\em Physical Review Letters}, 103(18):184501, 2009.

\bibitem{miljkovic2013jumping}
Nenad Miljkovic, Ryan Enright, Youngsuk Nam, Ken Lopez, Nicholas Dou, Jean
  Sack, and Evelyn~N Wang.
\newblock Jumping-droplet-enhanced condensation on scalable superhydrophobic
  nanostructured surfaces.
\newblock {\em Nano Letters}, 13(1):179--187, 2013.

\bibitem{baba2020dropwise}
Soumei Baba, Kenichiro Sawada, Kohsuke Tanaka, and Atsushi Okamoto.
\newblock Dropwise condensation on a hierarchical nanopillar structured
  surface.
\newblock {\em Langmuir}, 36(34):10033--10042, 2020.

\bibitem{das2021filmwise}
Chayan Das, Rohit Gupta, Saikat Halder, Amitava Datta, and Ranjan Ganguly.
\newblock Filmwise condensation from humid air on a vertical superhydrophilic
  surface: Explicit roles of the humidity ratio difference and the degree of
  subcooling.
\newblock {\em Journal of Heat Transfer}, 143(6):061601, 2021.

\bibitem{wen2018three}
Rongfu Wen, Shanshan Xu, Xuehu Ma, Yung-Cheng Lee, and Ronggui Yang.
\newblock Three-dimensional superhydrophobic nanowire networks for enhancing
  condensation heat transfer.
\newblock {\em Joule}, 2(2):269--279, 2018.

\bibitem{liu2021coalescence}
Chuntian Liu, Meirong Zhao, Yelong Zheng, Luya Cheng, Jiale Zhang, and Clarence
  Augustine~TH Tee.
\newblock Coalescence-induced droplet jumping.
\newblock {\em Langmuir}, 37(3):983--1000, 2021.

\bibitem{wang2020density}
Rui Wang, Feifei Wu, Dandan Xing, Fanfei Yu, and Xuefeng Gao.
\newblock Density maximization of one-step electrodeposited copper nanocones
  and dropwise condensation heat-transfer performance evaluation.
\newblock {\em ACS Applied Materials \& Interfaces}, 12(21):24512--24520, 2020.

\bibitem{gerasopoulos2018effects}
Konstantinos Gerasopoulos, William~L Luedeman, Emre Olceroglu, Matthew
  McCarthy, and Jason~J Benkoski.
\newblock Effects of engineered wettability on the efficiency of dew
  collection.
\newblock {\em ACS Applied Materials \& Interfaces}, 10(4):4066--4076, 2018.

\bibitem{nusselt1916surface}
W~Nusselt.
\newblock The surface condensation of water vapour.
\newblock {\em Zeitschrift Des Vereines Deutscher Ingenieure}, 60:541--546,
  1916.

\bibitem{beysens2022physics}
Daniel Beysens.
\newblock {\em The Physics of Dew, Breath Figures and Dropwise Condensation},
  volume 994.
\newblock Springer, 2022.

\bibitem{zhang2021analytical}
Wei Zhang, Suilin Wang, and Mu~Lianbo.
\newblock Analytical modeling for vapor condensation in the presence of
  noncondensable gas and experimental validation.
\newblock {\em Journal of Heat Transfer}, 143(1):011601, 2021.

\bibitem{luo2016new}
Xisheng Luo, Yu~Fan, Fenghua Qin, Huaqiao Gui, and Jianguo Liu.
\newblock A new model for the processes of droplet condensation and evaporation
  on solid surface.
\newblock {\em International Journal of Heat and Mass Transfer}, 100:208--214,
  2016.

\bibitem{liu2020heat}
Weihong Liu and Xiang Ling.
\newblock Heat transfer model based on diffusion layer theory for dropwise
  condensation with high non-condensable gas.
\newblock {\em AIP Advances}, 10(12):125305, 2020.

\bibitem{vedder1969aluminum+}
W~Vedder and DA~Vermilyea.
\newblock Aluminum+ water reaction.
\newblock {\em Transactions of the Faraday Society}, 65:561--584, 1969.

\bibitem{peng2018all}
Chaoyi Peng, Zhuyang Chen, and Manish~K Tiwari.
\newblock All-organic superhydrophobic coatings with mechanochemical robustness
  and liquid impalement resistance.
\newblock {\em Nature Materials}, 17(4):355--360, 2018.

\end{thebibliography}


\begin{thebibliography}{10}

\bibitem{feng2012condensate}
Jie Feng, Yichuan Pang, Zhaoqian Qin, Ruiyuan Ma, and Shuhuai Yao.
\newblock Why condensate drops can spontaneously move away on some
  superhydrophobic surfaces but not on others.
\newblock {\em ACS Applied Materials \& Interfaces}, 4(12):6618--6625, 2012.

\bibitem{feng2012factors}
Jie Feng, Zhaoqian Qin, and Shuhuai Yao.
\newblock Factors affecting the spontaneous motion of condensate drops on
  superhydrophobic copper surfaces.
\newblock {\em Langmuir}, 28(14):6067--6075, 2012.

\bibitem{rykaczewski2013multimode}
Konrad Rykaczewski, Adam~T Paxson, Sushant Anand, Xuemei Chen, Zuankai Wang,
  and Kripa~K Varanasi.
\newblock Multimode multidrop serial coalescence effects during condensation on
  hierarchical superhydrophobic surfaces.
\newblock {\em Langmuir}, 29(3):881--891, 2013.

\bibitem{zhang2017enhanced}
Peng Zhang, Yota Maeda, Fengyong Lv, Yasuyuki Takata, and Daniel Orejon.
\newblock Enhanced coalescence-induced droplet-jumping on nanostructured
  superhydrophobic surfaces in the absence of microstructures.
\newblock {\em ACS Applied Materials \& Interfaces}, 9(40):35391--35403, 2017.

\bibitem{beysens2022physics}
Daniel Beysens.
\newblock {\em The Physics of Dew, Breath Figures and Dropwise Condensation},
  volume 994.
\newblock Springer, 2022.

\bibitem{das2021filmwise}
Chayan Das, Rohit Gupta, Saikat Halder, Amitava Datta, and Ranjan Ganguly.
\newblock Filmwise condensation from humid air on a vertical superhydrophilic
  surface: Explicit roles of the humidity ratio difference and the degree of
  subcooling.
\newblock {\em Journal of Heat Transfer}, 143(6):061601, 2021.

\bibitem{kim2011}
Sunwoo Kim and Kwang~J. Kim.
\newblock Dropwise condensation modeling suitable for superhydrophobic
  surfaces.
\newblock {\em Journal of Heat Transfer}, 133(8), 05 2011.

\bibitem{carey2020liquid}
Van~P Carey.
\newblock {\em Liquid-vapor phase-change phenomena: An introduction to the
  thermophysics of vaporization and condensation processes in heat transfer
  equipment}.
\newblock CRC Press, 2020.

\bibitem{tanasawa1991advances}
Ichiro Tanasawa.
\newblock Advances in condensation heat transfer.
\newblock In {\em Advances in Heat Transfer}, volume~21, pages 55--139.
  Elsevier, 1991.

\bibitem{luo2016new}
Xisheng Luo, Yu~Fan, Fenghua Qin, Huaqiao Gui, and Jianguo Liu.
\newblock A new model for the processes of droplet condensation and evaporation
  on solid surface.
\newblock {\em International Journal of Heat and Mass Transfer}, 100:208--214,
  2016.

\bibitem{zheng2018modeling}
Shaofei Zheng, Ferdinand Eimann, Christian Philipp, Tobias Fieback, and Ulrich
  Gross.
\newblock Modeling of heat and mass transfer for dropwise condensation of moist
  air and the experimental validation.
\newblock {\em International Journal of Heat and Mass Transfer}, 120:879--894,
  2018.

\bibitem{aoki2003knudsen}
Kazuo Aoki, Claude Bardos, and Shigeru Takata.
\newblock Knudsen layer for gas mixtures.
\newblock {\em Journal of Statistical Physics}, 112(3):629--655, 2003.

\bibitem{young1993condensation}
JB~Young.
\newblock The condensation and evaporation of liquid droplets at arbitrary
  knudsen number in the presence of an inert gas.
\newblock {\em International Journal of Heat and Mass Transfer},
  36(11):2941--2956, 1993.

\bibitem{jacobson1999fundamentals}
Mark~Z Jacobson and Mark~Z Jacobson.
\newblock {\em Fundamentals of atmospheric modeling}.
\newblock Cambridge university press, 1999.

\bibitem{wen2019falling}
Rongfu Wen, Xingdong Zhou, Benli Peng, Zhong Lan, Ronggui Yang, and Xuehu Ma.
\newblock Falling-droplet-enhanced filmwise condensation in the presence of
  non-condensable gas.
\newblock {\em International Journal of Heat and Mass Transfer}, 140:173--186,
  2019.

\bibitem{zhang2021analytical}
Wei Zhang, Suilin Wang, and Mu~Lianbo.
\newblock Analytical modeling for vapor condensation in the presence of
  noncondensable gas and experimental validation.
\newblock {\em Journal of Heat Transfer}, 143(1):011601, 2021.

\bibitem{nusselt1916surface}
W~Nusselt.
\newblock The surface condensation of water vapour.
\newblock {\em Zeitschrift Des Vereines Deutscher Ingenieure}, 60:541--546,
  1916.

\bibitem{uchida1964evaluation}
Hi~Uchida, A~Oyama, and Y\_ Togo.
\newblock Evaluation of post-incident cooling systems of light water power
  reactors.
\newblock Technical report, Tokyo Univ., 1964.

\end{thebibliography}

\end{document}


\maketitle

\section*{Section S1: Microscale condensation behavior on superhydrophobic surfaces}

Figure S1 shows the zoomed-in view of the condensate droplets appearing at different instants of time on the SHB surfaces of copper and aluminum, so as to discern the coalescence, sliding and jumping behaviors of milimetric and submillimetric condensate droplets. During the coalescence of micro-drops on the SHB copper surface, the merged drop either jumped from the substrate or remained attached to the substrate, depending on whether the excess surface energy due to the coalescence could overcome the work of the solid-liquid adhesion or not \cite{feng2012condensate, feng2012factors}. Figure S1a shows the coalescence of two similar-sized drops, which caused out-of-plane jumping without disturbing any other neighbouring drops. On the contrary, the coalescence of asymmetric drops has been found to result in the spontaneous jumping of merged drops with an in-plane component of velocity. This in turn leads to further coalescence with neighbouring drops, and results in multidrop jumping \cite{rykaczewski2013multimode}, as shown in Fig. S1b. The interspace distance between the top microstructures of aluminum substrate is larger than the copper substrate (Fig. 1), which increases the diffusion of water vapor to the crevices of the roughness features, allowing more condensate to form within the microfeatures \cite{feng2012factors, zhang2017enhanced}. This causes a higher solid-liquid adhesion on the aluminum SHB surface due to pinning, and the condensate drops do not jump upon coalescence. In Fig. S1c, during the coalescence of unequal-sized drops, the smaller droplets were completely depinned from their initial location and merged with the bigger droplet. When multiple drops were present in close proximity to the three phase contact line of the merging drops, the triggering of a single coalescence could also result in multidrop merging (see Fig. S1d).

\section*{Section S2: Influence of non-condensable gases during condensation}
\subsection*{Nucleation}

Condensation initiates with nucleation, i.e., the formation of thermodynamically stable liquid embryos on a substrate at a few particular sites depending on the surface roughness, wettability of a substrate, supersaturation ratio, etc. The nucleation rate of a substrate is defined as the probability of that liquid embryo continuing to grow without evaporation. The heterogeneous nucleation rate per unit area is given by \cite{beysens2022physics}
\begin{equation}
    \frac{dN}{dt}=\frac{R_0}{3}Ae^{-\frac{W}{k_BT_s}}
\end{equation}
where $R_0$ is the radius of the water molecule, $R_0 \approx 1.375\times10^{-10}$ $m$, $A$ is the Arrhenius prefactor, $W$ is the total energy required for the formation of a liquid nuclei and $T_s$ is the substrate temperature. The expression of $W$ is as follows,
\begin{equation*}
    W=\frac{\beta^3}{27\alpha^2}(2-3cos\theta+cos\theta^3)
\end{equation*}
where $\theta$ is the contact angle, $\alpha=k_BT_sln(S)$ and $\beta=4\pi\sigma_{lv}\big(\frac{3v_m}{4\pi}\big)^{2/3}$

Here $S$ is called as supersaturation ratio and it is the ratio of partial pressure of the vapor ($P_v$) to the saturated vapor pressure ($P_s$) corresponds to substrate temperature $T_s$ ($P_s$). $\sigma_{lv}$ is the surface tension of water, $v_m$ is the volume occupied by a liquid molecule. The magnitude of $v_m$ for water is $v_m \approx 3\times10^{-29}$ $m^3$.

The magnitude of The Arrhenius prefactor is evaluated by the multiplication of three factors
\begin{equation}
    A=c_0Zf^*
\end{equation}
Here, $c_0$ is the volumetric concentration of the available nucleation sites and expressed as
\begin{equation}
    c_0=\frac{P_m}{k_BT_s}
\end{equation}
where $P_m$ is the total pressure of the humid air. The term $Z$ is called as the Zeldovich parameter and expressed as 
\begin{equation}
    Z=\frac{3\alpha^2}{4\beta^2}\bigg(\frac{\beta}{\pi k_B T_s}\bigg)^{1/2}
\end{equation}
$f^*$ is called as attachment frequency and mathematically expressed as
\begin{equation}
    f^*=(48\pi^2v_m)^{1/3}\alpha_mSx_sDn^{*1/3}
\end{equation}
where $D$ is the diffusion coefficient of water vapor molecules in air, $\alpha_m$ is the monomer sticking coefficient and approximated as 1 \citep{beysens2022physics}, $x_s$ is the saturated water molecule concentration in air and calculated from ideal gas equation 
\begin{equation}
    x_s=\frac{P_s}{k_BT_s}
\end{equation}
$n^*$ is the number of molecules occupied by the nuclei having a critical radius of $r_{cr}$ and given by,
\begin{equation}
    n^*=\bigg(\frac{2\beta}{3\alpha}\bigg)^{2/3}
\end{equation}

\subsection*{Temperature drop}

The major driving parameters for humid air condensation are degree of subcooling ($\Delta T$) and humidity ratio difference ($\Delta \omega$) \cite{das2021filmwise}. A theoretical analysis was performed for a wide range of non-condensable gas (NCG) concentrations in order to better understand the role of NCG during condensation. This section compares the individual contribution of temperature drop due to different thermal resistances at a moderate subcooling of 10 $K$.       
\subsubsection*{Dropwise condensation}
Figure 4B shows a schematic representation of the thermal resistance network model of a single droplet for dropwise condensation in the presence of NCG. The schematic identifies three regions  pertinent to the thermal resistance: the promoter layer coating region, the condensed drop/liquid region, and the vapor region. The vapor region includes an interfacial region called Knudsen layer and a surrounded diffusion layer region. 

The total temperature difference ($\Delta T$) between the cold surface and bulk vapor region for a single droplet can be expressed as,
\begin{equation}
\Delta T= \Delta T_{coat}+\Delta T_{drop}+\Delta T_{vapor}
\label{eq:dT}
\end{equation}
where $\Delta T_{coat}$ is the temperature drop across the promoter layer coating due to conduction and mathematically expressed as,
\begin{equation}
\Delta T_{coat}=\frac{q_d \delta _{coat}}{k_{coat}\pi r_d^2 sin^2\theta}
\label{eq:dT_coat}
\end{equation} 
where $q_d$ is the heat transfer through a drop of radius $r_d$, $\delta _{coat}$ is the thickness of the coating, $k_{coat}$ is the thermal conductivity of the coating, and $\theta$ is the equilibrium contact angle of a sessile water droplet on the coated surface. The temperature drop across the condensate drop ($\Delta T_{drop}$) due to conduction is given by \cite{kim2011},
\begin{equation}
\Delta T_{drop}=\frac{q_d \theta}{4\pi r_d k_{l} sin\theta}
\label{eq:dT_drop}
\end{equation}  
where $k_{l}$ is the thermal conductivity of the condensate liquid. The temperature difference in the vapor region ($\Delta T_{vapor}$) is calculated based on the bulk vapor conditions. 

In the case of pure vapor condensation, the temperature at the outer interface of the Knudsen layer ($T_i$) maintained as saturation temperature ($T_{sat}$) corresponding to the ambient pressure and vapor region keeps on an equilibrium state. $\Delta T_{vapor}$ is defined as the sum of two temperature drops such as $\Delta T_c$ and $\Delta T_{int}$. Wherein, the temperature drop due to curvature resistance ($\Delta T_c$) is expressed as,
\begin{equation}
\Delta T_c=\frac{2T_{sat}\sigma_{lv}}{h_{lv} r_d \rho_l}=\frac{r_{cr}}{r_d}\Delta T
\label{eq:dT_c}
\end{equation}
where $h_{lv}$ is the latent heat for vapor-liquid phase change, $\rho_l$ is the density of the condensate liquid and $r_{cr}$ is the critical radius for nucleation. The interfacial temperature drop ($\Delta T_{int}$) is expressed as,
\begin{equation}
\Delta T_{int}=\frac{q_d}{h_{int}2\pi r_d^2(1-cos\theta)}
\label{eq:dT_int}
\end{equation} 
In the case of the condensation of pure vapor $h_{int}$ is calculated from \cite{carey2020liquid},
\begin{equation}
h_{int}=\frac{2\epsilon}{2-\epsilon}\frac{1}{\sqrt{2\pi R_{v} T_{sat}}}\frac{h_{lv}^2}{v _{v} T_{sat}}\bigg(1-\frac{p_vv_v}{2h_{lv}}\bigg)
\label{eq:h_int_vap}
\end{equation}
where $\epsilon$ is the accommodation co-efficient, $R_v$ is the gas constant for the water vapor, $p_v$ saturated vapor pressure corresponding to $T_{sat}$, and $v_v$ is the specific volume. The magnitude of $\epsilon$ is dependent on the condensation conditions and is approximated as 0.02 for pure vapor conditions \cite{tanasawa1991advances}. The heat transfer rate across a single droplet of radius $r_d$ during pure vapor condensation can be calculated using Eq. \ref{eq:dT} after substituting Eq. \ref{eq:dT_coat} - \ref{eq:dT_int},
\begin{equation}
(q_d)_{pure\:vapor}=\frac{\Delta T \pi r^2(1-\frac{r_{cr}}{r_d})}{\bigg[\frac{\delta _{coat}}{k_{coat}sin^2\theta}+\frac{r_d\theta}{4k_l sin\theta}+\frac{1}{2h_{int}(1-cos\theta)}\bigg]}
\label{eq:q_d}
\end{equation} 

In the case of condensation in the presence of NCG, a diffusion layer forms outside the interfacial region and a temperature gradient of vapor exists in this region during condensation. Therefore, interfacial temperature outside the Knudsen layer can only be predicted from the coupled solution of mass diffusion and energy conservation laws. Hence the heat transfer rate, the interface temperature between the coating and drop ($T_{coat}$), the surface temperature of the drop ($T_d$), and the Knudsen region outer layer temperature ($T_i$) for each condensing drops are calculated numerically by simultaneous solution of mass and energy conservation laws \cite{luo2016new,zheng2018modeling}. 

In the Knudsen layer, the kinetic theory of gases governs the transfer of vapor molecules to the liquid drop interface. The effect of molecular collision in the Knudsen layer is negligible, and the thickness of the Knudsen layer is in the order of mean free path ($\lambda_m$) of the vapor-NCG mixture which can be mathematically expressed as \cite{aoki2003knudsen},
\begin{equation}
\lambda_m=\frac{2\mu_m\sqrt{R_mT_m}}{P_m}
\end{equation}
where $\mu_m$, $R_m$, $T_m$, $P_m$ are the dynamic viscosity, gas constant, temperature and pressure of the vapor-NCG mixture respectively. The outer radius of the Knudsen layer ($r_i$) is expressed as 
\begin{equation}
 r_i=r_d+\beta\lambda_m   
\end{equation} 
where $r_d$ is the condensed drop radius, $\beta$ is a constant and the best fit value for $\beta$ is 0.75 \cite{young1993condensation}. 

The heat transfer rate across the droplet and coating region is governed by the Fourier law of heat conduction. The mass transfer rate through the coating and the droplet region is given by,
\begin{equation}
   \dot{m}=\frac{T_d-T_s}{h_{lv}\bigg[\frac{\delta _{coat}}{k_{coat}sin^2\theta}+\frac{r_d\theta}{4k_l sin\theta}\bigg]}
   \label{eqn:m1}
\end{equation}
The mass transfer rate across the Knudsen layer region is calculated from the kinetic theory of gases \cite{zheng2018modeling},  
\begin{equation}
\dot{m}=\frac{4\pi r_d^2 r_i^2 (1-cos\theta)}{2r_i^2-\alpha r_d^2}\Bigg(\frac{\alpha \rho_{v,i}R_vT_i}{\sqrt{2\pi R_vT_i}}-\frac{\alpha \rho_{v,d}R_vT_d}{\sqrt{2\pi R_vT_d}}\Bigg)
    \label{eqn:m2}
\end{equation}
where $\rho_{v,i}$, $\rho_{v,d}$ are the water vapor density at the Knudsen interface and liquid drop surface, $R_v$ is the gas constant of the water vapor. The magnitude of $\rho_{v,i}$, $\rho_{v,d}$ are calculated from the ideal gas equation as,
\begin{equation}
    \rho_{v,i}=\frac{P_i}{R_vT_i}
\end{equation}
\begin{equation}
    \rho_{v,d}=\frac{P_d}{R_vT_d}
\end{equation}
where the vapor pressure at the Knudsen interface is given by \cite{jacobson1999fundamentals} 
\begin{equation}
    P_i=611.2 \: exp \Big[6816\Big(\frac{1}{273.15}- \frac{1}{T_i}\Big)+ 5.1309\:ln\Big(\frac{273.15}{T_i}\Big)\Big]
\end{equation}
and the vapor pressure at the curved interface of the droplet surface is derived from the Kelvin equation as follows \cite{luo2016new}, 
\begin{equation}
    P_d=exp\Bigg(\frac{2\sigma_{lv}}{\rho_lR_vT_dr_d}\Bigg)\Bigg[611.2 \: exp \Big[6816\Big(\frac{1}{273.15}- \frac{1}{T_d}\Big)+ 5.1309\:ln\Big(\frac{273.15}{T_d}\Big)\Big]\Bigg]
\end{equation}
In Eq. \ref{eqn:m2}, the variable $\alpha$ is condensation coefficient and the magnitude of $\alpha$ vary with the vapor conditions. The magnitude of $\alpha$ at higher NCG condition ($\sim 97\%$) is approximated by validating the numerical model with \textit{Zheng et al.} experiments \cite{zheng2018modeling} and the lower NCG condition ($\sim 5\%$) is approximated by validating the numerical model with \textit{Wen et al.} experiments \cite{wen2019falling}.
The value of $\alpha$ is found to be 0.0003 and 0.0075 for the NCG mass fraction of $\sim$97\% and $\sim$5\% respectively. The $\alpha$ value for other NCG fractions is approximated from the exponential fitting curve between the computed $\alpha$ magnitudes for the NCG mass fraction of 97\% and 5\%. 
The mass flux across the diffusion layer is calculated based on the Fick's law of diffusion given by,
\begin{equation}
\dot{m}=2\pi r_i(1-cos\theta)D_m(\rho_{v,m}-\rho_{v,i})
    \label{eqn:m3}
\end{equation}
where $\rho_{v,m}$ is the density of water vapor outside the diffusion layer based on the vapor pressure at free stream. The diffusion co-efficient ($D_m$) of vapor-NCG mixture is calculated from the Chapman-Enskog theory \cite{jacobson1999fundamentals}, 
\begin{equation}
    D_m=2.11\times10^{-5}\bigg(\frac{T_m}{273.15}\bigg)^{1.94}\bigg(\frac{1}{P_m}\bigg)
\end{equation}
Based on the mass conservation, Eq. \ref{eqn:m1}, \ref{eqn:m2}, \ref{eqn:m3} can be solved simultaneously. These equations consists of three unknown variables: $\dot{m}$, $T_d$ and $T_i$. A Matlab script has been used for the computation of the unknown parameters by iterative solution of the equations Eq. \ref{eqn:m1}, \ref{eqn:m2}, \ref{eqn:m3}. The heat transfer rate per droplet in presence of NCG is given by,
\begin{equation}
    (q_d)_{NCG}=\dot{m}h_{lv}
\end{equation}
 
Figures 4D compare the magnitude of temperature drop across the droplet and vapor region for a condensate drop of radius 100 $nm$ - 1 $mm$ during the condensation of pure vapor and the vapor-NCG mixture. The figures are plotted for a moderate subcooling of 10 $^\circ C$ with $\delta_{coat}$= 1 $\mu m$, $k_{coat}$= 0.2 $W/mK$, $\theta$= 90 $^\circ$, $k_l$= 0.6 $W/mK$, $h_{lv}$= 2257 $kJ/kg$, $\sigma_{lv}$= 0.072 $N/m$, $\rho_l$= 997 $kg/m^3$, $P_v$=101.325 $kPa$, $T_{sat}$=373.15 $K$, $v_v$= 1.672 $m^3/kg$ \cite{kim2011}.  

\subsubsection*{Filmwise condensation}
The total temperature difference ($\Delta T$) between the cold surface and bulk vapor region during filmwise condensation can be expressed as,
\begin{equation}
\Delta T= \Delta T_{film}+\Delta T_{vapor}
\label{eq:dT_f}
\end{equation}
The temperature drop across the condensate film ($\Delta T_{film}$) due to conduction is given by,
\begin{equation}
\Delta T_{film}=\frac{q''\delta_f}{k_{l}}
\label{eq:dT_drop_f}
\end{equation}
where $q''$ is the average heat flux, $\delta_f$ is the average condensate film thickness approximated as $\sim$25 $\mu m$ \cite{zhang2021analytical}. 

The average heat flux during pure vapor condensation is calculated from the Nusselt film condensation theory \cite{nusselt1916surface},
\begin{equation}
    q''_{pure\:vapor}=0.943\Bigg(\frac{g\rho_l(\rho_l-\rho_v)k_l^3h_{lv}'}{\mu_l\Delta T L}\Bigg)^{0.25}\Delta T
\end{equation}
where $g$ is the acceleration due to gravity, $h_{lv}'=h_{lv}+C_p\Delta T$ in which $C_p$ is heat capacity at a constant pressure, $\mu_l$ is the viscocity of the condensate film, $L$ is length of the substrate.

The average heat flux for the condensation of vapor-NCG mixture is approximated from the Uchida correlation \cite{uchida1964evaluation},
\begin{equation}
q''_{NCG}=U_{Uchida}\Delta T= 380\bigg(\frac{\omega_v}{1-\omega_v}\bigg)^{0.7}\Delta T
\label{eq:q_ncg}
\end{equation}  
where $U_{Uchida}$ is the overall condensation heat transfer co-efficient for the vapor-NCG mixture environment and $\omega_v$ is the percentage of vapor content in the mixture. Figure 4E is plotted using this analysis for a subcooling of 10 $^\circ C$.

\bibliographystyle{unsrt}
\bibliography{SupplementaryFiles/supplementary}

\newpage
\begin{figure}[t]
    \centering
    \includegraphics[width=\linewidth]{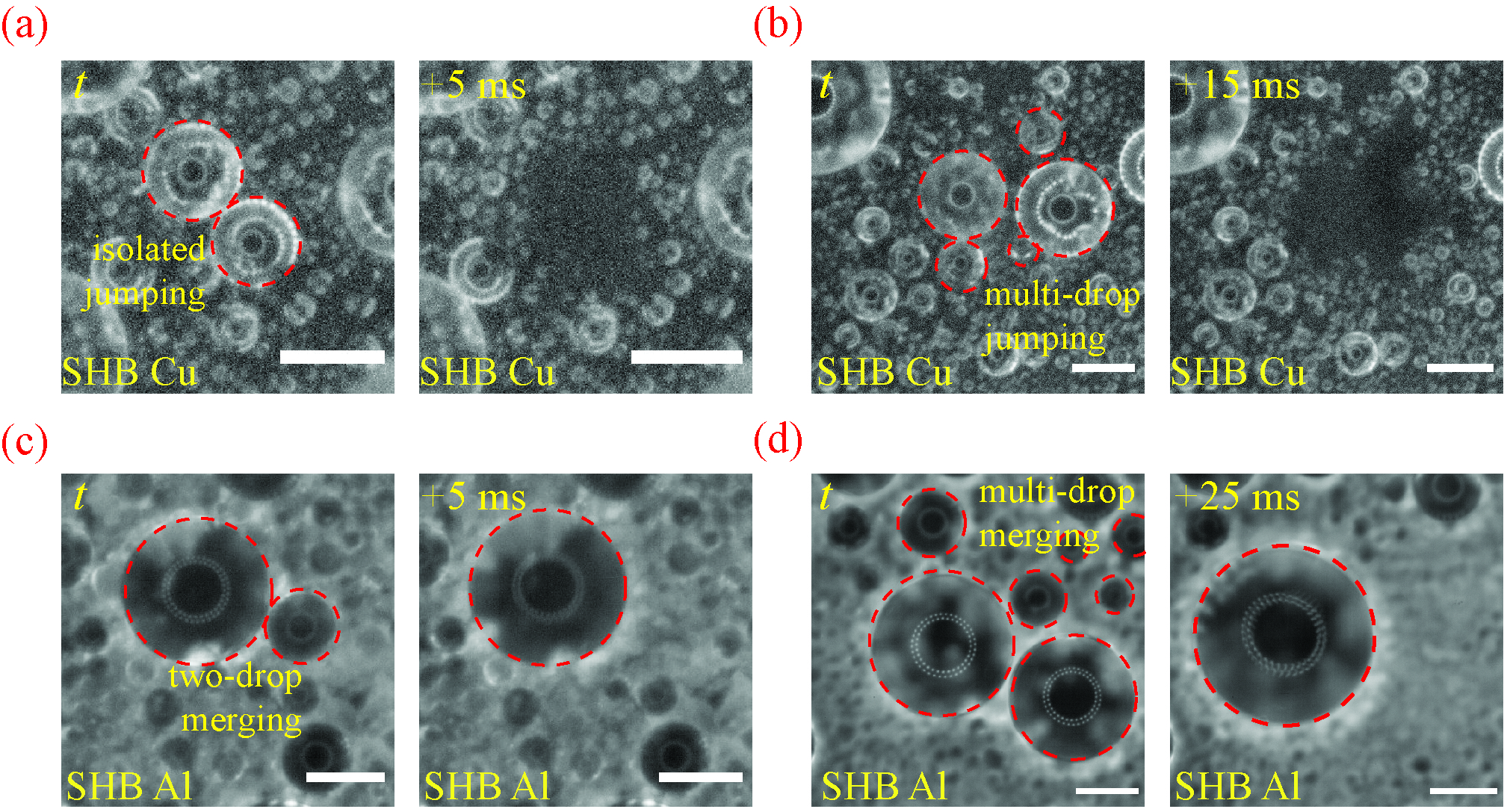}
    \caption{Microscale condensation on superhydrophobic surfaces. Coalescence-induced jumping (CIDJ) of (a) two-isolated drops and (b) multidrops on a SHB copper substrate. Drop merging of (c) two drops (d) multi drops during the coalescence of condensate drops on a SHB aluminum substrate. The environmental condition was $T_{env}$= 30 $^\circ C$ and $RH$= 60 \%. Images were recorded using a high-speed camera at 200 $fps$ using a zoom lens (Navitar-Zoom 6000). The resolution of the image was 1.06 $\mu m$/$pixel$. The scale bar is 100 $\mu m$ for (a,b,c) and 200 $\mu m$ for (d).}    
    \label{fig:microdrop}
\end{figure}

\begin{figure}[t]
    \centering
    \includegraphics[width=\linewidth]{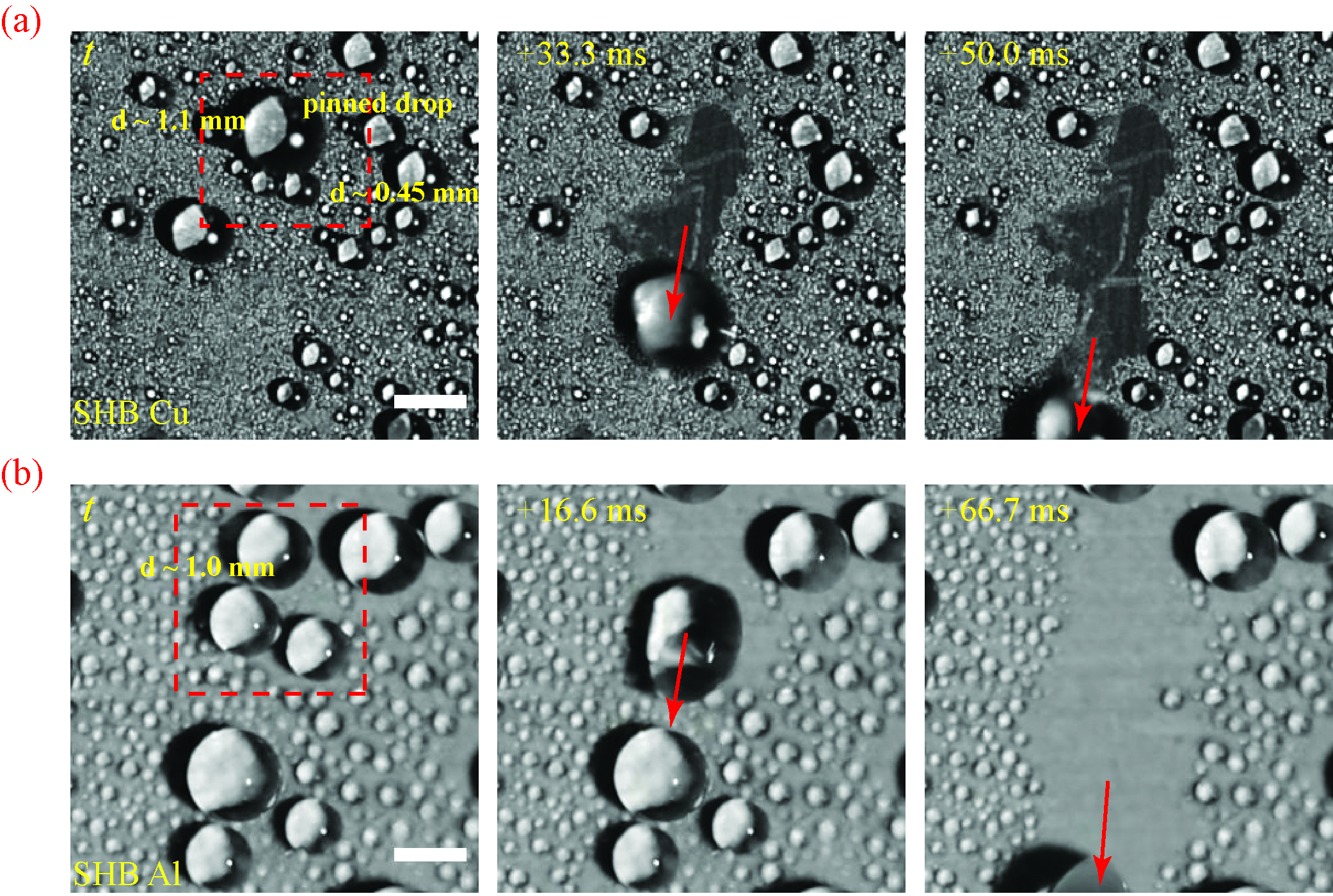}
    \caption{Condensate drainage from superhydrophobic surfaces. The sweeping mechanisms of (a) SHB copper substrate and (b) SHB aluminum substrate at $T_{env}$= 30 $^\circ C$ and $RH$= 60 \%. The scale bar is 1 $mm$. }    
    \label{fig:sweep}
\end{figure}

\begin{figure}[t]
    \centering
    \includegraphics[width=\linewidth]{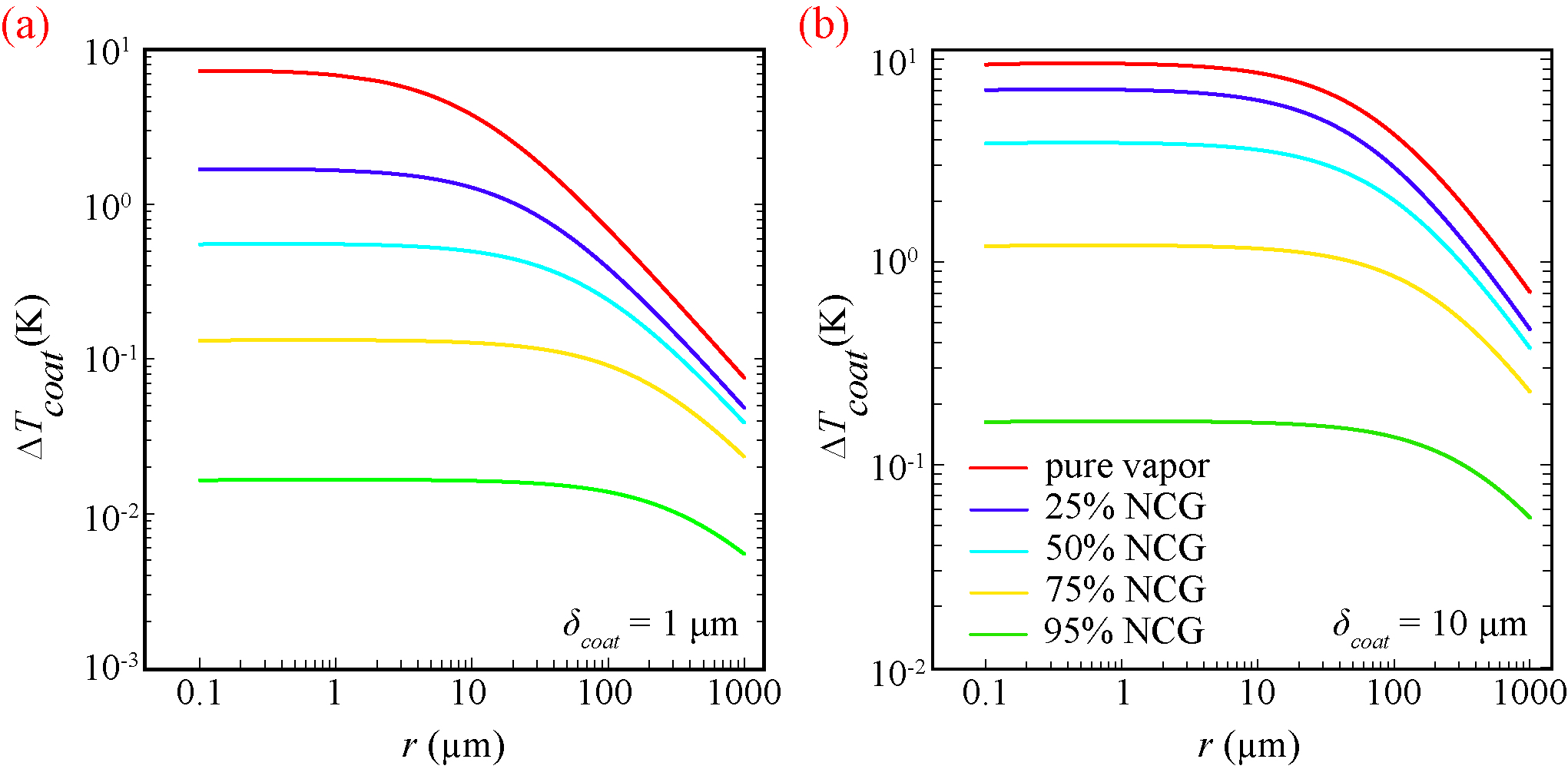}
    \caption{The temperature drop across the hydrophobic coating. The effect of NCG concentrations on the temperature drop across the hydrophobic promoter layer coating ($\Delta T_{coat}$) of thickness (a) 1 $\mu m$, (b) 10 $\mu m$.}
    \label{fig:coat}
\end{figure}

\begin{figure}[t]
    \centering
    \includegraphics[width=\linewidth]{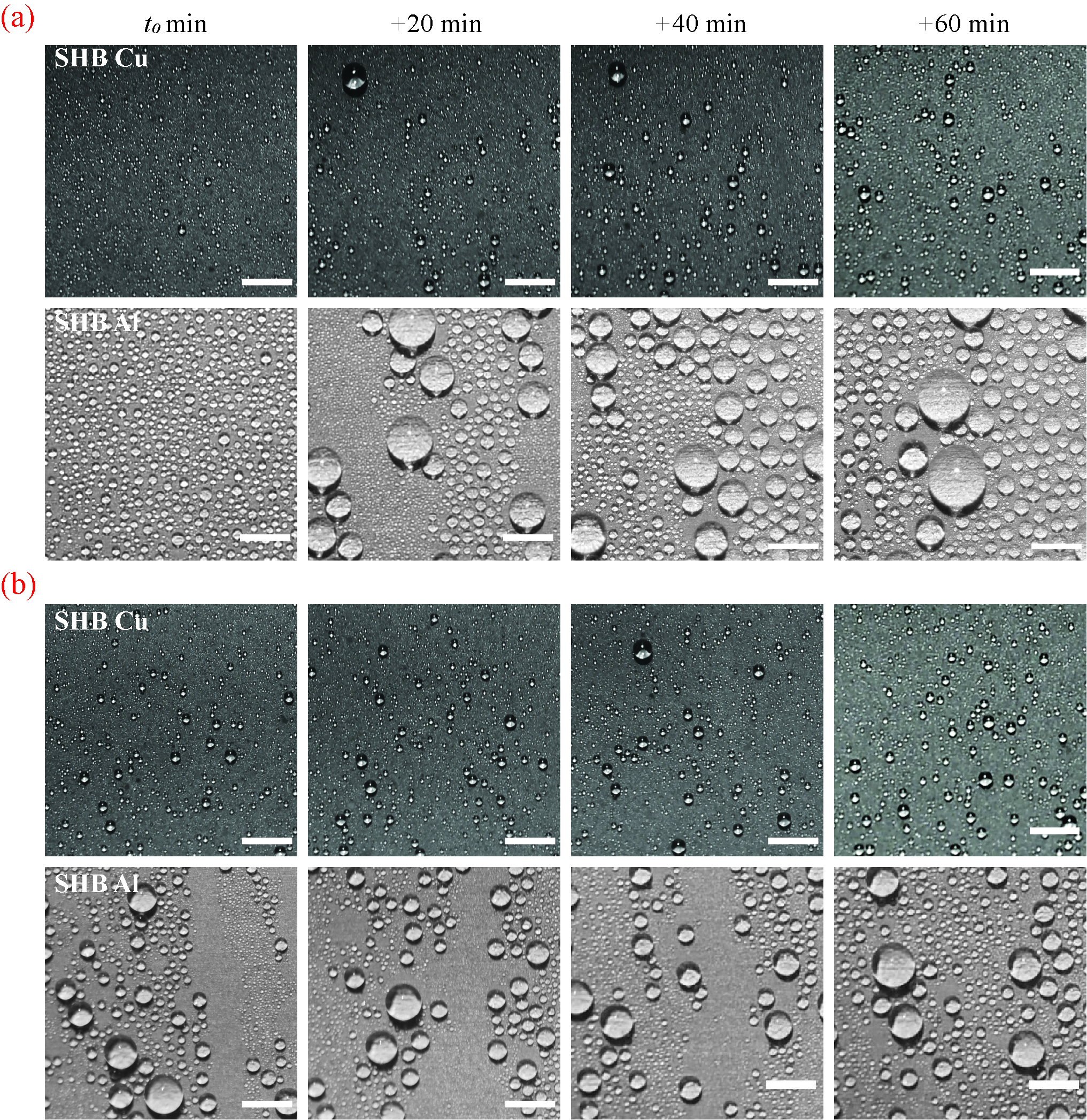}
    \caption{The observation of dropwise condensation at macro-scale. The time-lapse snapshots of condensation behavior on SHB copper and aluminum substrates at (a) $T_{env}$= 20 $^\circ C$ and $RH$= 75 \%, (b) $T_{env}$= 45 $^\circ C$ and $RH$= 90 \%. $t_0$ is the time at which the surface temperature reaches the set value. The scale bar is 2 $mm$. }    
    \label{fig:snaps}
\end{figure} 
 
\begin{figure}[t]
    \centering
    \includegraphics[width=0.8\linewidth]{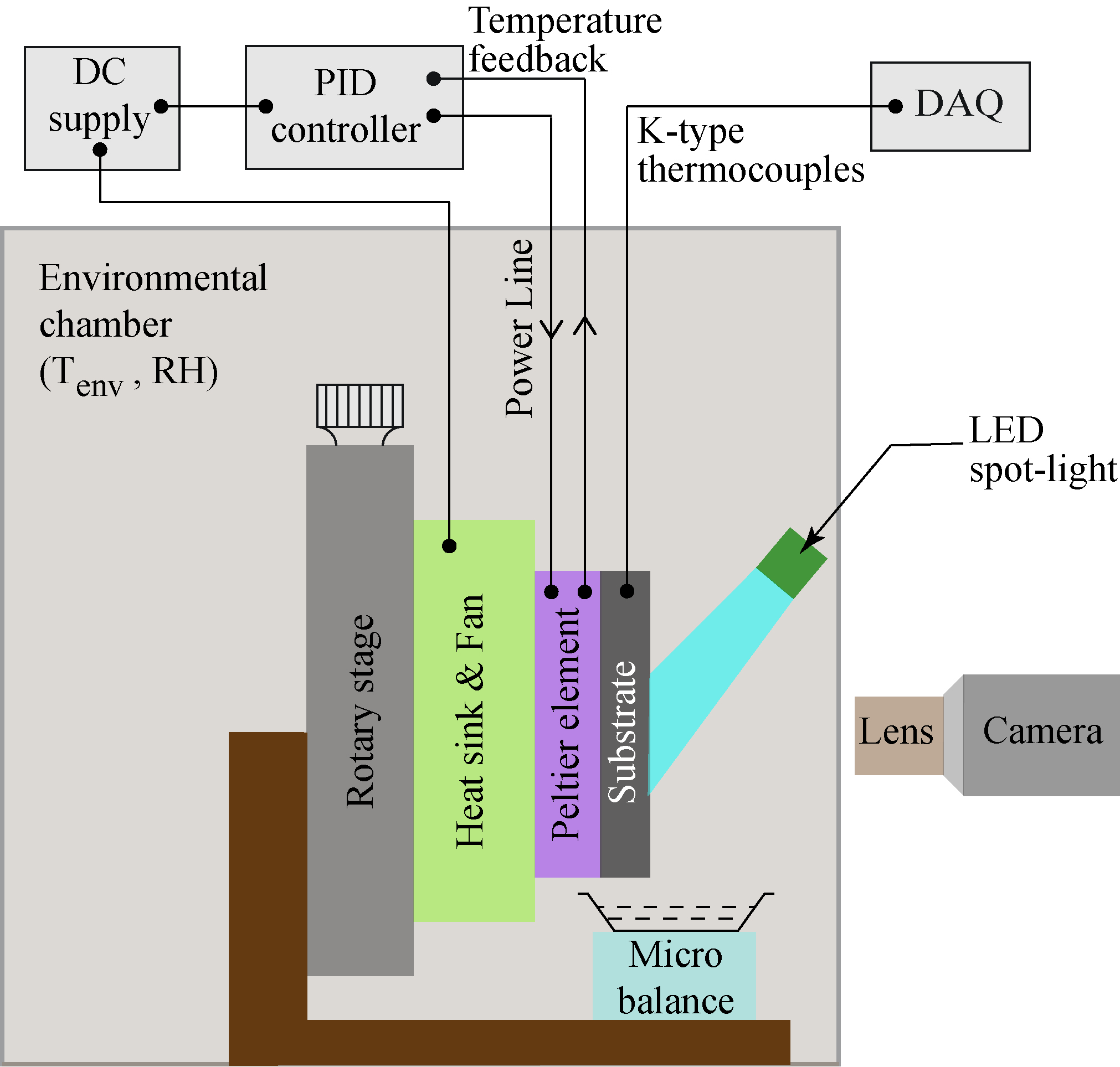}
    \caption{Schematics of the experimental setup. The condensation experiments were performed in a controlled environmental chamber. The temperature of the substrate was maintained at a constant temperature using a Peltier element.}
    \label{fig:setup}
\end{figure}


\medskip